\documentclass[%
 reprint,
 amsmath,amssymb,
 aps,
]{revtex4-1}
\pdfoutput=1
\usepackage{amsthm}
\usepackage{graphicx}
\usepackage{dcolumn}
\usepackage{bm}
\usepackage{extarrows}
\usepackage{float}
\usepackage[roman]{complexity}

\begin{document}

\title{A quantum algorithm for solving eigenproblem of the Laplacian matrix of a fully connected weighted graph}

\author{Hai-Ling Liu $^{1,2}$}

\author{Su-Juan Qin $^{1}$}
\email{qsujuan@bupt.edu.cn}

\author{Lin-Chun Wan $^{1}$}

\author{Chao-Hua Yu $^{3}$}

\author{Shi-Jie Pan $^{1}$}

\author{Fei Gao $^{1}$}
\email{gaof@bupt.edu.cn}

\author{Qiao-Yan Wen $^{1}$}

\affiliation{$^{1}$ State Key Laboratory of Networking and Switching Technology, Beijing University of Posts and Telecommunications, Beijing 100876, China}
\affiliation{$^{2}$ State Key Laboratory of Cryptology, P.O. Box 5159, Beijing, 100878, China}
\affiliation{$^{3}$ School of Information Management, Jiangxi University of Finance and Economics, Nanchang 330032, China}
\date{\today}

\begin{abstract}
Solving eigenproblem of the Laplacian matrix of a fully connected weighted graph has wide applications in data science, machine learning, and image processing, etc. However, this is very challenging because it involves expensive matrix operations. Here, we propose an efficient quantum algorithm to solve it based on a assumption that the element of each vertex and its norms can be effectively accessed via a quantum random access memory data structure. Specifically, we adopt the optimal Hamiltonian simulation technique based on the block-encoding framework to implement the quantum simulation of the Laplacian matrix. Then, the eigenvalues and eigenvectors of the Laplacian matrix are extracted by the quantum phase estimation algorithm. The core of our entire algorithm is to construct the block-encoding of the Laplacian matrix. To achieve this, we propose in detail how to construct the block-encodings of operators containing the information of the weight matrix and the degree matrix respectively, and further obtain the block-encoding of the Laplacian matrix. Compared with its classical counterpart, our algorithm has a polynomial speedup on the number of vertices and an exponential speedup on the dimension of each vertex. We also show that our algorithm can be extended to solve the eigenproblem of symmetric (non-symmetric) normalized Laplacian matrix.
\end{abstract}

\pacs{Valid PACS appear here}
\maketitle


\section{Introduction}

 Quantum computing has exhibited potential acceleration advantages over classical computing by exploiting the unique properties of supposition and entanglement in quantum mechanics in solving certain problems, such as factoring integers \cite{S1994}, unstructured database searching \cite{LKG1997}, solving equations \cite{HHL2009,WY2018,LWWP2021}, regression \cite{NDS2012,YGW2019}, dimensionality reduction \cite{SMP2014,IL2016,PWL2020,P2022}, anomaly detection \cite{L2018,G2021} and neural network \cite{ASZL2021}. Overviews on quantum algorithms can be seen in Refs.\cite{MA2016,BJW2017}.

 In the era of big data, graph learning \cite{XKSAL2021} has attracted considerable attention owing to its wide applications in data science, machine learning, and image processing, etc. In the process of dealing with problems related to graph learning, such as graph networks \cite{ZCZC2020,ADSM2021}, image processing \cite{RYEMMP2017,GG2009}, and reinforcement learning \cite{WZQJ2021}, it is necessary to solve the eigenproblem of the Laplacian matrix of a fully connected weighted graph to avoid dropping crucial nonlocal information. In general, the Laplacian matrix $L$ is given by the difference between the degree matrix $D$ and the weight matrix $W$ $(L=D-W)$. For a fully connected weighted graph, $W$ is a dense matrix, thus $L$ is also a dense matrix. However, solving the eigenproblem of a dense matrix $L$ is very challenging because it involves expensive matrix operations. Therefore, it is imperative for us to design an efficient algorithm to solve this problem.

 Fuelled by the success of quantum algorithms, some scholars proposed to solve the eigenproblem of $L$ with quantum algorithm. Kerenidis et al. \cite{KIL2021} proposed a quantum algorithm for solving the eigenproblem of the symmetric normalized Laplacian matrix, and successfully applied it to the spectral clustering algorithm. However, as the authors point out, their algorithm cannot efficiently access $D$ and the norm of the rows vectors of $W$. This makes their algorithm cannot be directly used to solve the eigenproblem of $L$ of a fully connected weighted graph. Subsequently, Li et al. \cite{2022} designed a quantum algorithm to solve the eigenproblem of $L$ and also applied it to accelerate the spectral clustering algorithm. However, their quantum algorithm can only handle the sparse matrix $L$.

In $2016$, Huang et al. proposed a quantum Laplacian eigenmap algorithm, which has a significant speedup compared with its classical counterparts \cite{HYX2016}. However, this algorithm relies on a strong assumption that the classically stored information of $W$ and $D$ can be superposition accessed by a quantum random access memory \cite{GVSM2008}. In fact, the data usually given is the vertex set of a graph in most scenarios \cite{ZCZC2020,ADSM2021,RYEMMP2017,GG2009,CG1997,V2007}, $W$ and $D$ need to be obtained by complex computations. This makes the classical complexity of implementing this strong assumption theoretically would exceed the complexity of the quantum algorithm itself. In short, the existing quantum algorithm cannot efficiently solve the eigenproblem of $L$ of a fully connected weighted graph.

In this paper, we design an efficient quantum algorithm to solve eigenproblem of the Laplacian matrix $L$ of a fully connected weighted graph without the above strong assumptions. The starting point of our quantum algorithm was the assumption of having superposition access to the element of each vertex and its norms. This makes our proposed algorithm more suitable for practical scenarios. Specifically, we adopt the optimal Hamiltonian simulation technique based on the block-encoding framework \cite{LGCI2019,GSYLW2019,CSGJ2018} to implement the quantum simulation of $L$, which reduce the algorithm's dependence on simulation error. Then we employ the quantum phase estimation algorithm \cite{NC2002} to extract the eigenvalues and eigenvectors of $L$. The core of our entire algorithm is to construct the block-encoding of $L$. To achieve it, we design the specific controlled unitary operators to prepare the quantum states to construct the block-encodings of operators containing the information of $W$ and $D$ respectively, and further obtain the block-encoding of $L$. Compared with its classical counterpart, our algorithm has a polynomial speedup on the number of vertices and an exponential speedup on the dimension of each vertex. In particular, our algorithm can also be used to solve eigenproblem of $W$, which is also of great significance \cite{CWWLF2008,GLZA2021,LYWHG2020,ZTLZC2020}. We also show that our algorithm can be extended to solve the eigenproblem of symmetric (non-symmetric) normalized Laplacian matrix.

The remainder of the paper is organized as follows. In Sec.~\ref{Sec:Review}, we give a brief overview of the graph Laplacian matrix. In Sec.~\ref{Sec:Quantum}, we propose a quantum algorithm to solve eigenproblem of the Laplacian matrix of a fully connected weighted graph. In Sec.~\ref{sec:generalization}, we generalized the algorithm to solve eigenproblem of symmetric (non-symmetric) normalized Laplacian matrix. In Sec.~\ref{Sec:Discussion}, we give some discussions. Finally, we present our conclusion in Sec.~\ref{Sec:Conclusion}.

\section{Review of the graph Laplacian matrix}
\label{Sec:Review}

Given a weighted undirected graph $G=(V,E)$ with the vertex set $V=\{\mathbf{x}_{i}|\mathbf{x}_{i}\in R^{m}\}_{i=1}^{n}$ and the edge set $E$. The weight matrix $W\in R^{n\times n}$ of the graph $G$ is defined as follows:
\begin{equation}
 w_{ij}=w_{ji}=\left\{
 \begin{aligned}
&\geq 0,& i\neq j,\\
 &=0, &i=j.\\
 \end{aligned}
 \right.
\end{equation}
Specifically, the element $w_{ij}>0$ represents that vertex $\mathbf{x}_{i}$ is connected to vertex $\mathbf{x}_{j}$, otherwise $w_{ij}=0$. For $w_{ij},i\neq j$, we take the Gaussian similarity function $w_{ij}=\exp(-\lambda\|\mathbf{x}_{i}-\mathbf{x}_{j}\|^{2})$ as an example which has a wide range of applications, where $\lambda>0$ is any given real number \cite{Be2003,KMMH2017}. It is worth noting that we can also choose other forms of nonlinear similar functions \cite{CG1997,V2007}, such as sigmoid and cosine similar function.

Next the degree matrix $D$ of the graph $G$ is defined as a diagonal matrix $D={\rm diag}(d_{ii})\in R^{n\times n}$, where $d_{ii}=\sum_{j=1}^{n}w_{ij}$.

Given $W$ and $D$, the graph Laplacian matrix $L$ is defined as
\begin{equation}
L=D-W\in R^{n\times n},
\end{equation}
where $L$ is a symmetric positive semi-definite matrix. In addition, there are two common normalized Laplacian matrices as follows:
\begin{equation}
\begin{aligned}
L_{s}&=D^{-\frac{1}{2}}LD^{-\frac{1}{2}}=I-D^{-\frac{1}{2}}WD^{-\frac{1}{2}},\\
L_{r}&=D^{-1}L=I-D^{-1}W,
\end{aligned}
\end{equation}
where $L_{s}$ $(L_{r})$ is a symmetric (non-symmetric) matrix. For more information about $L$, see Refs.\cite{CG1997,V2007}.

Solving eigenproblem of $L$ of a fully connected weighted graph has wide applications in graph networks \cite{ZCZC2020,ADSM2021}, image processing \cite{RYEMMP2017,GG2009}, and reinforcement learning \cite{WZQJ2021}, etc. For a fully connected weighted graph, $W$ is a dense matrix, thus $L$ is also a dense matrix. And the multiplicity of the zero eigenvalue of $L$ is equal to $1$ and the eigenvector corresponding to zero eigenvalue is $\mathbf{1}=(1,\cdots,1)^{T}\in R^{n}$ \cite{V2007}. It provides no extra information. Therefore, we consider extracting $1\leq d\leq(n-1)$ non-zero eigenvalues and the corresponding eigenvectors of $L$, and its classical complexity is $O(mn^{2}+dn^{3})$ \cite{SZR2014}. This is quite time consuming when the size of the vertex set is large. Therefore, it is imperative for us to design an efficient algorithm to solve it.

\section{A quantum algorithm for solving eigenproblem of the Laplacian matrix of a fully connected weighted graph}
\label{Sec:Quantum}

To design an efficient quantum algorithm to solve eigenproblem of $L$ of a fully connected weighted graph,
 we adopt the optimal Hamiltonian simulation technology based on the block-encoding framework \cite{LGCI2019,GSYLW2019,CSGJ2018} to realize the quantum simulation of $L$ and then perform quantum phase estimation algorithm \cite{NC2002} to extract the eigenvalues and eigenvectors of $L$. The core of our entire algorithm is to construct the block-encoding of $L$. To achieve it,
 we construct the block-encodings of operators containing the information of $W$ and $D$ respectively, and further obtain the block-encoding of $L$.

The entire section consists of five subroutines: we review the optimal Hamiltonian simulation technology based on the block-encoding framework in Sec.~\ref{Sec:R}, quantum algorithms for preparing the quantum states to construct the block-encodings of operators containing the information of $W$ and $D$ in Sec.~\ref{Sec:W} and Sec.~\ref{Sec:D}, respectively, a quantum algorithm for obtaining the block-encoding of $L$ to implement the quantum simulation of $L$ in Sec.~\ref{Sec:I}, and a quantum algorithm to extract the eigeninformation of $L$ in Sec.~\ref{Sec:Extract}. For convenience, we define the base of the logarithm function as $2$, which can be abbreviated as $\log x$.

Assume that the vertex set $V=\{\mathbf{x}_{i}|\mathbf{x}_{i}\in R^{m}\}_{i=1}^{n}$ is stored in a quantum random access memory (QRAM) data structure \cite{WLZZ2018}, i.e., the element $x_{ij}$ of each vertex $\mathbf{x}_{i}$ is stored in the $i$th leaf of the binary tree, and the internal node of the tree stores the modulo sum of the elements in the subtree rooted in it. Then there exists a quantum algorithm that can perform the following map with $\varepsilon_{x}$-precision in $O[\poly\log(mn/\varepsilon_{x})]$ time:
\begin{equation}
U|i\rangle|0\rangle\longrightarrow|i\rangle|\mathbf{x}_{i}\rangle=|i\rangle\frac{1}{\|\mathbf{x}_{i}\|}\sum_{j=1}^{m}x_{ij}|j\rangle.
\end{equation}
In addition, this structure can perform the unitary operator $O$ in time $O[\poly\log(mn)]$:
\begin{equation}
O|i\rangle|0\rangle\longrightarrow|i\rangle|\|\mathbf{x}_{i}\|\rangle.
\end{equation}

 This data structure has also been successfully applied to quantum data compression \cite{YGLJ2019}, quantum linear systems with displacement structures \cite{WCP2021} and so on.

\subsection{Review of the optimal Hamiltonian simulation technology based on the block-encoding framework}
\label{Sec:R}

In this section, we review the optimal Hamiltonian simulation technology based on the block-encoding framework \cite{LGCI2019,GSYLW2019,CSGJ2018}. We first give the framework of block-encoding.

$\mathbf{Definition 1}$ (Block-encoding). Assume that $A$ is an $s$-qubits operator, $\alpha$, $\epsilon_{A}\in \mathbb{R}^{+}$, and $a\in \mathbb{N}$, then we say that the $(s+a)$-qubits unitary $U$ is an $(\alpha,a,\epsilon_{A})$ block-encoding of $A$ if it satisfies
\begin{equation}
\|A-\alpha(\langle0|^{\otimes a}\otimes I)U(|0\rangle^{\otimes a}\otimes I)\|\leq\epsilon_{A}.
\end{equation}

Meanwhile, Low and Chuang \cite{LGCI2019} also proposed a block-encoding framework of a purified density operator, as follows:

$\mathbf{Lemma 1}$ (Block-encoding of density operators).
Suppose that $\rho$ is an $s$-qubits density operator and $G$ is an $(a+s)$-qubits unitary operator that acts  on the input state $|0\rangle^{\otimes a}|0\rangle^{\otimes s}$ prepares a purification $|0\rangle^{\otimes a}|0\rangle^{\otimes s}\mapsto |\rho\rangle$, s.t. ${\rm Tr}_{a}(|\rho\rangle\langle\rho|)=\rho$. Then $(G^{\dagger}\otimes I_{s})(I_{a}\otimes SWAP_{s})(G\otimes I_{s})$ is an $(1,a+s,0)$-block-encoding of $\rho$.

Subsequently, Andr\'{a}s Gily\'{e}n et al. \cite{GSYLW2019} proposed to implement a block-encoding of a linear combination of block-encoded operators. It is shown as follows:

$\mathbf{Definition 2}$ (State preparation pair).
Let $\mathbf{y}\in \mathbb{C}^{m}$ and $\|\mathbf{y}\|_{1}\leq\beta$. The pair of unitaries $(P_{L},P_{R})$ is called an $(\beta,b,\varepsilon_{y})$-state-preparation-pair if $P_{L}|0\rangle^{\otimes b}=\sum_{j=1}^{2^{b}}c_{j}|j\rangle$ and $P_{R}|0\rangle^{\otimes b}=\sum_{j=1}^{2^{b}}d_{j}|j\rangle$ such that $\sum_{j=1}^{m}|\beta(c_{j}^{\ast}d_{j})-y_{j}|\leq\varepsilon_{y}$ and for all $j\in m+1,\cdots,2^{b}$, we have $c_{j}^{\ast}d_{j}=0$.

$\mathbf{Lemma 2}$ (Linear combination of block-encoded matrices).
Let $A=\sum_{j=1}^{m}y_{j}A_{j}$ be an $s$-qubits operator and $\varepsilon_{A}\in \mathbb{R}^{+}$. Assume that $(P_{L},P_{R})$ is an $(\beta,b,\varepsilon_{y})$-state-preparation-pair for $\mathbf{y}\in \mathbb{C}^{m}$,
$W =\sum_{j=1}^{m}|j\rangle\langle j|\otimes U_{j}+((I-\sum_{j=1}^{m}|j\rangle\langle j|)\otimes I_{a}\otimes I_{s})$ is an $(s+a+b)$-qubits unitary operator such that for all $j=1,\cdots,m$, we have that $U_{j}$ is an $(\alpha,a,\varepsilon_{A})$-block-encoding of $A_{j}$. Then we can implement an $(\alpha\beta,a+b,\alpha\varepsilon_{y}+\alpha\beta\varepsilon_{A})$-block-encoding
of $A$, with a single use of $W$, $P_{R}$ and $P_{L}^{\dag}$.

Based on the above block-encoding framework, the optimal Hamiltonian simulation technique is proposed as follows:

$\mathbf{Theorem 1}$ (Optimal Block-Hamiltonian simulation). Suppose that $U$ is an $(\alpha,a,\varepsilon/|2t|)$-block-encoding of the Hamiltonian $H$. Then we can implement an $\varepsilon$-precise Hamiltonian simulation unitary $V$ which is an $(1,a+2,\varepsilon)$-block-encoding of $\exp(itH)$, with $O(|\alpha t|+\log(1/\varepsilon)/\log\log(1/\varepsilon))$ uses of controlled-$U$ or its inverse and with
$O(a|\alpha t|+a\log(1/\varepsilon)/\log\log(1/\varepsilon))$ two-qubit gates.

The core of this technique is to construct the block-encoding of an operator. To realize the quantum simulation of $L=D-W$, next we will design quantum algorithms to construct the block-encodings of operators containing the information of $W$ and $D$, respectively.

\subsection{Prepare the quantum state to construct the block-encoding of an operator containing the information of $W$}
\label{Sec:W}
We know that the elements of $W$ are
\begin{equation}
\begin{aligned}
w_{ij}&=\exp(-\lambda\|\mathbf{x}_{i}-\mathbf{x}_{j}\|^{2})\\
&=\exp[-\lambda(\|\mathbf{x}_{i}\|^{2}+\|\mathbf{x}_{j}\|^{2})]\exp(2\lambda\mathbf{x}_{i}\cdot\mathbf{x}_{j}),i\neq j,
\end{aligned}
\end{equation}
where $i,j=1,2,\cdots,n$.

Due to $\exp[-\lambda(\|\mathbf{x}_{i}\|^{2}+\|\mathbf{x}_{j}\|^{2})]$ is a scalar that depends on the size of $\|\mathbf{x}_{i}\|, i=1,\cdots,n$, we take the Taylor expansion of  $\exp(2\lambda\mathbf{x}_{i}\cdot\mathbf{x}_{j})$ to get
\begin{equation}
\exp(2\lambda\mathbf{x}_{i}\cdot\mathbf{x}_{j})=\sum_{k=0}^{\infty}\frac{(2\lambda)^{k}}{k!}(\mathbf{x}_{i}\cdot\mathbf{x}_{j})^{k}.
\end{equation}
According to Taylor's theorem \cite{KM1998}, by keeping only the low-order terms of the
Taylor expansion, we can get a finite-dimensional approximated of $\exp(2\lambda\mathbf{x}_{i}\cdot\mathbf{x}_{j})$, i.e.,
 \begin{equation}
 \exp(2\lambda\mathbf{x}_{i}\cdot\mathbf{x}_{j})=\sum_{k=0}^{p}\frac{(2\lambda)^{k}}{k!}(\mathbf{x}_{i}\cdot\mathbf{x}_{j})^{k}.
 \end{equation}
 A detailed analysis of the errors of low order approximation with Taylor expansion, see Ref.\cite{LPC2011}. 

 Combine Eq.$(7)$ with Eq.$(9)$, we can obtain
 \begin{equation}
 w_{ij}=\exp[-\lambda(\|\mathbf{x}_{i}\|^{2}+\|\mathbf{x}_{j}\|^{2})]\sum_{k=0}^{p}a_{k}(\mathbf{x}_{i}\cdot\mathbf{x}_{j})^{k},
 \end{equation}
 where $a_{k}=(2\lambda)^{k}/k!$, $k=0,1,\cdots,p$. It is worth noting that other forms of nonlinear similarity functions can also obtain corresponding low-order approximations by Taylor expansion, whose forms are similar to $w_{ij}$, such as sigmoid and cosine similarity functions.

Note that when $\|\mathbf{x}_{i}\|=1$, according to Eq.$(10)$, we have
\begin{equation}
\exp[-\lambda(\|\mathbf{x}_{i}\|^{2}+\|\mathbf{x}_{j}\|^{2})]=\exp(-2\lambda).
\end{equation}
Once $\lambda$ is given, $\exp(-2\lambda)$ is a constant that can be absorbed into $a_{k}$, namely $\tilde{a}_{k}=[\exp(-2\lambda)(2\lambda)^{k}]/k!$. However, when $\|\mathbf{x}_{i}\|\neq1$, its value will affects the value of $a_{k}$, which result in $\exp[-\lambda(\|\mathbf{x}_{i}\|^{2}+\|\mathbf{x}_{j}\|^{2})]$ not being absorbed into $a_{k}$.

 Therefore, next we will design the corresponding quantum algorithms to construct the block-encodings of operators containing the information of $W$ in two cases, as follows:

\textbf{(I.1) A quantum algorithm in the case of $\|\mathbf{x}_{i}\|=1,i=1,\cdots,n$}

The specific steps of the quantum algorithm are as follows.

$\rm{(I.1)}$ Prepare the quantum state
\begin{equation}
\frac{1}{\sqrt{n}}\sum_{i=1}^{n}|i\rangle_{1}\frac{1}{\sqrt{\tilde{a}}}\sum_{k=0}^{p}\sqrt{\tilde{a}_{k}}|k\rangle_{2}|0^{\log m}\rangle^{\otimes p}_{3},\tilde{a}=\sum_{k=0}^{p}\tilde{a}_{k}.
\end{equation}
Here, assume that the vector $\mathbf{\tilde{a}}=\frac{1}{\sqrt{\tilde{a}}}(\sqrt{\tilde{a}_{0}},\cdots,\sqrt{\tilde{a}_{p}})^{T}\in R^{p+1}$ is stored in a QRAM data structure \cite{WLZZ2018}. Then there is a quantum algorithm that can generate an $\varepsilon_{\tilde{a}}$-approximation of  $|\mathbf{\tilde{a}}\rangle=\frac{1}{\sqrt{\tilde{a}}}\sum_{k=0}^{p}\sqrt{\tilde{a}_{k}}|k\rangle$ with gate complexity $O\{\poly\log[(p+1)/\varepsilon_{\tilde{a}}]\}$.

$\rm{(I.2)}$ Apply the controlled unitary operator $R_{U}:=\sum_{k=0}^{p}|k\rangle_{2}\langle k|_{2}\otimes (I^{p-k}U^{k})_{1,3}$ to three registers, the system state will becomes
\begin{equation}
\begin{aligned}
&\frac{1}{\sqrt{n}}\sum_{i=1}^{n}|i\rangle_{1}\frac{1}{\sqrt{\tilde{a}}}\sum_{k=0}^{p}\sqrt{\tilde{a}_{k}}|k\rangle_{2}
|0^{\log m}\rangle^{\otimes p-k}_{3}|\mathbf{x}_{i}\rangle^{\otimes k}_{3}\\
&:=\frac{1}{\sqrt{n}}\sum_{i=1}^{n}|i\rangle_{1}|\Phi(\mathbf{x}_{i})\rangle_{2,3}:=|\Phi\rangle,
\end{aligned}
\end{equation}
where
\begin{equation}
\begin{aligned}
&|\Phi(\mathbf{x}_{i})\rangle_{2,3}=\frac{1}{\sqrt{\tilde{a}}}(\sqrt{\tilde{a}_{0}}|0\rangle|0^{\log m}\rangle^{\otimes p}\\
&+\sqrt{\tilde{a}_{1}}|1\rangle|0^{\log m}\rangle^{\otimes p-1}|\mathbf{x}_{i}\rangle+\cdots+\sqrt{\tilde{a}_{p}}|p\rangle|\mathbf{x}_{i}\rangle^{\otimes p})_{2,3}.
\end{aligned}
\end{equation}

\begin{figure}
\centering
	\includegraphics[width=8.5cm]{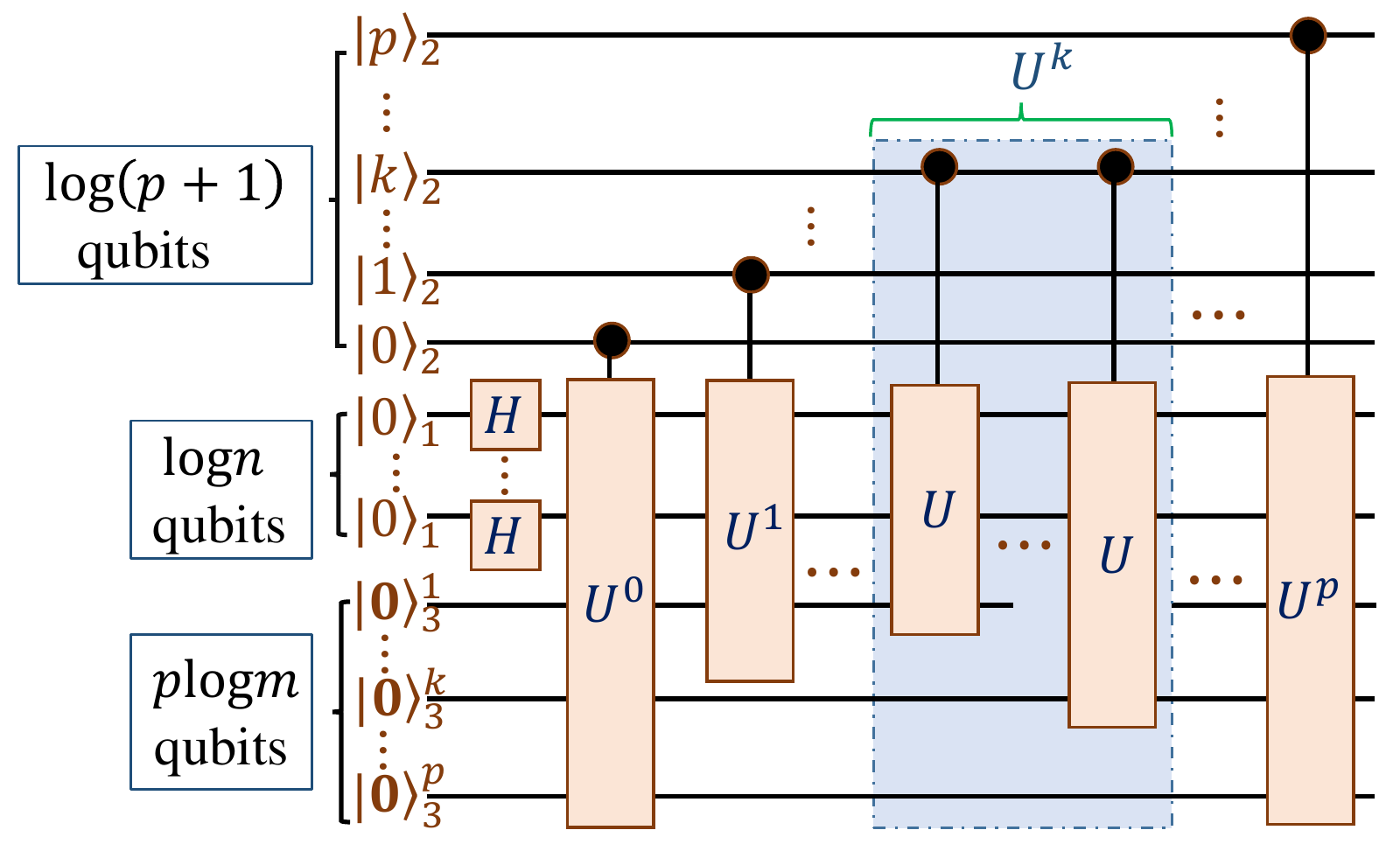}
	\caption{The quantum circuit of the controlled unitary operator $R_{U}$. Here the subscripts $1,2,3$ represent the index of the registers. The third register has $p\log m$ qubits which comes from $p$ sub-registers with $\log m$ qubits, namely, $|\mathbf{0}\rangle_{3}^{k}=|0^{\log m}\rangle_{3}^{k}(k=1,\cdots,p)$ and $H$ denotes Hadamard gate, and $U^{k},k=0,1,\cdots,p,$ represent $k$ consecutive executions of the unitary operator $U$. }
	\label{FIG:1}
\end{figure}

The quantum circuit of implementing $R_{U}$ is shown in Fig.~\ref{FIG:1}.
Note that
\begin{equation}
\begin{aligned}
\langle\Phi(\mathbf{x}_{i})|\Phi(\mathbf{x}_{j})\rangle&=[\tilde{a}_{0}+\tilde{a}_{1}\langle\mathbf{x}_{i}|\mathbf{x}_{j}\rangle+\cdots+\tilde{a}_{p}(\langle\mathbf{x}_{i}|\mathbf{x}_{j}\rangle)^{p}]/\tilde{a}\\
&=w_{ij}/\tilde{a}.
\end{aligned}
\end{equation}

$\rm{(I.3)}$ Take partial trace for the second and third registers, we have
\begin{equation}
\begin{aligned}
\rho_{0}:={\rm Tr}(|\Phi\rangle\langle\Phi|)_{2,3}&=\frac{1}{n}\sum_{i=1}^{n}\langle\Phi(\mathbf{x}_{i})|\Phi(\mathbf{x}_{j})\rangle|i\rangle_{1}\langle j|_{1}\\
&=(W+\tilde{a}I)/n\tilde{a}.
\label{16}
\end{aligned}
\end{equation}

According to $\mathbf{Lemma1}$, we can design an $(a_{0}+s_{0})$-qubits unitary operator $G_{0}$ which implement
$G_{0}|0\rangle^{\otimes a_{0}}|0\rangle^{\otimes s_{0}}\mapsto |\Phi\rangle$, s.t.
${\rm Tr}_{2,3}(|\Phi\rangle\langle\Phi|)=\rho_{0}$, where $a_{0}=p\log m+\log(p+1)$, $s_{0}=\log n$.
Then $V_{0}:=(G_{0}^{\dagger}\otimes I_{1})(I_{2,3}\otimes SWAP_{1})(G_{0}\otimes I_{1})$ is an
$(1,a_{0}+s_{0},2\epsilon_{0})$-block-encoding of $\rho_{0}$, where $\epsilon_{0}$ is the error that produces the state $|\Phi\rangle$. See $\textbf{(I.2)}$ for a detailed analysis about $\epsilon_{0}$. For convenience, we use $\hat{\cdot}$ to represent the estimated value caused by the quantum algorithm in the following sections.

\textbf{(I.2) The complexty of algorithm in the case of $\|\mathbf{x}_{i}\|=1,i=1,\cdots,n$ }

In step $\rm{(I.1)}$, the complexity is $O\{\poly\log[n(p+1)/\varepsilon_{\tilde{a}}]\}$, which is derived from $\log n$ Hadamard gates generating the first register and $O\{\poly\log[(p+1)/\varepsilon_{\tilde{a}}]\}$ with gate complexity generating the second register.

In step $\rm{(I.2)}$, according to the quantum circuit of $R_{U}$ in Fig.~\ref{FIG:1}, it need $O(p)$ calls of the unitary operator $U$. Thus the complexity of $R_{U}$ is $O(p\cdot \poly\log(mn/\varepsilon_{x}))$. Next we analyze the error that produces the state $|\Phi\rangle$ as follows:
\begin{equation}
\begin{aligned}
&\||\hat{\Phi}\rangle-|\Phi\rangle\|_{2}\\
&=\|\frac{1}{\sqrt{n}}\sum_{i=1}^{n}|i\rangle|\hat{\Phi}(\mathbf{x}_{i})\rangle-\frac{1}{\sqrt{n}}\sum_{i=1}^{n}|i\rangle|\Phi(\mathbf{x}_{i})\rangle\|_{2}\\
&\leq\frac{1}{\sqrt{n}}\sum_{i=1}^{n}\||\hat{\Phi}(\mathbf{x}_{i})\rangle-|\Phi(\mathbf{x}_{i})\rangle\|_{2}\\
&\leq\frac{1}{\sqrt{n}}\sum_{i=1}^{n}[\varepsilon_{\tilde{a}}+(1+2+\cdots+p)\varepsilon_{x}]\\
&=\sqrt{n}[\varepsilon_{\tilde{a}}+(1+2+\cdots+p)\varepsilon_{x}],
\end{aligned}
\label{17}
\end{equation}
where the final inequality is proved in Appendix~\ref{Sec:analysis}.

 To make the error to generate $|\Phi\rangle$ is $\epsilon_{0}$, we choose $\varepsilon_{\tilde{a}}=\varepsilon_{x}$, then
\begin{equation}
\||\hat{\Phi}\rangle-|\Phi\rangle\|_{2}\leq\sqrt{n}p^{2}\varepsilon_{x}:=\epsilon_{0}.
\end{equation}
We can get $1/\varepsilon_{x}=\sqrt{n}p^{2}/\epsilon_{0}$.

In short, the complexity is $O[p\cdot\poly\log(mn^{3/2}p^{2}/\epsilon_{0})]$.

\textbf{(II.2) A quantum algorithm in the case of $\|\mathbf{x}_{i}\|\neq1, i=1,\cdots,n$}

To facilitate the introduction of the algorithm, we first list the lemma required.

$\mathbf{Lemma 3}$ \cite{ZSS2017} (Quantum Multiply-Adder (QMA))
Let integers $\mathbf{a}$ and $\mathbf{b}$ be $m$-bit string. Then there is a quantum algorithm with $O[\poly(m)]$ single- and two-qubit gates can realize
\begin{equation}
\begin{aligned}
&|\mathbf{a}\rangle|\mathbf{b}\rangle\mapsto|\mathbf{a}\rangle|\mathbf{ab}\rangle,\\
&|\mathbf{a}\rangle|\mathbf{b}\rangle\mapsto|\mathbf{a}\rangle|\mathbf{a+b}\rangle.\\
\end{aligned}
\end{equation}
Note that for accuracy defined as  $\epsilon_{m}=2^{-m}$, the complexity of QMA is given by $O[\poly\log(1/\epsilon_{m})]$.

The specific algorithm proceeds as following steps:

$\rm{(II.1)}$ Prepare the quantum state
\begin{equation}
\begin{aligned}
&\frac{1}{\sqrt{n}}\sum_{i=1}^{n}|i\rangle_{1}\frac{1}{\sqrt{a}}\sum_{k=0}^{p}\sqrt{a_{k}}|k\rangle_{2}|0^{\log(mn)}\rangle^{\otimes
p}_{3}|0^{\log(mn)}\rangle^{\otimes 2}_{4}\\
&|0^{\log m}\rangle^{\otimes p}_{5},
\end{aligned}
\end{equation}
where $a=\sum_{k=0}^{p}a_{k}$ and the state $\frac{1}{\sqrt{a}}\sum_{k=0}^{p}\sqrt{a_{k}}|k\rangle$ can be prepared with precision $\varepsilon_{a}$ in the gate complexity of $O\{\poly\log[(p+1)/\varepsilon_{a}]\}$ which is similar to step$\rm{(I.1)}$.

\begin{figure}
\centering
	\includegraphics[width=8.5cm]{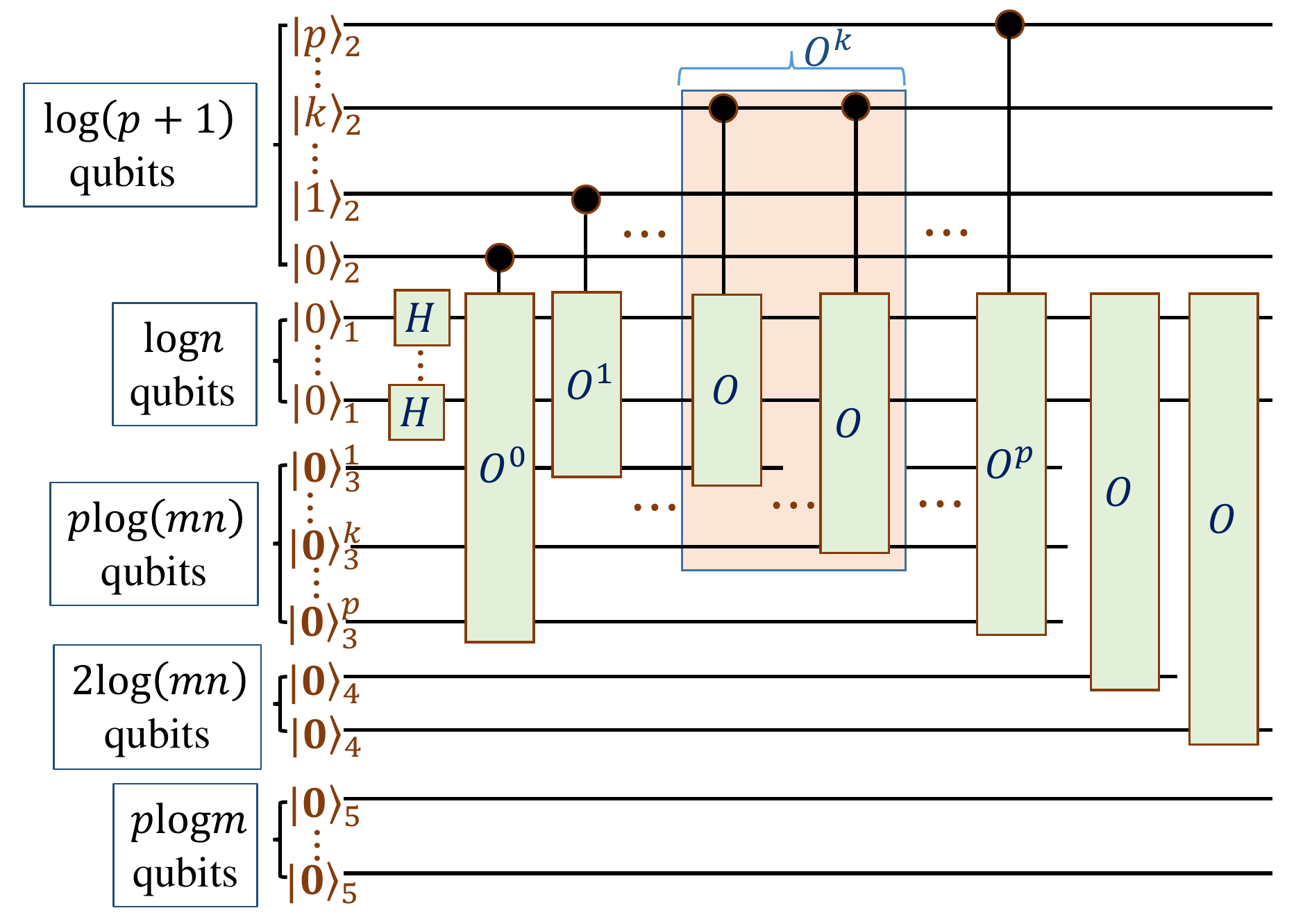}
	\caption{The quantum circuit of the controlled unitary operator $R_{O}$. The subscripts $1,\cdots,5$ represent the index of the registers. Here, each sub-register in the third and fourth registers has $\log(mn)$ qubits, i.e., $|\mathbf{0}\rangle=|0^{\log(mn)}\rangle$, and each sub-register in the fifth register has $\log m$ qubits, i.e., $|\mathbf{0}\rangle_{5}=|0^{\log m}\rangle_{5}$ and $H$ denotes Hadamard gate, and $O^{k},k=0,1,\cdots,p,$ represent $k$ consecutive executions of the unitary operator $O$.}
	\label{FIG:2}
\end{figure}

$\rm{(II.2)}$ Perform the controlled unitary operator $R_{O}:=\sum_{k=0}^{p}|k\rangle_{2}\langle k|_{2}\otimes (O^{k}I^{p-k})_{1,3}\otimes O^{2}_{1,4}\otimes I_{5}$ to get
\begin{equation}
\begin{aligned}
&\frac{1}{\sqrt{n}}\sum_{i=1}^{n}|i\rangle_{1}\frac{1}{\sqrt{a}}\sum_{k=0}^{p}\sqrt{a_{k}}|k\rangle_{2}|\|\mathbf{x}_{i}\|\rangle^{\otimes k}_{3}|0^{\log (mn)}\rangle^{\otimes p-k}_{3}\\
&|\|\mathbf{x}_{i}\|\rangle^{\otimes 2}_{4}|0^{\log m}\rangle^{\otimes p}_{5}.
\end{aligned}
\end{equation}
The quantum circuit of implementing $R_{O}$ in Fig.~\ref{FIG:2}.

$\rm{(II.3)}$ Apply $\mathbf{Lemma 3}$ to the third and fourth registers respectively, and undo the redundant registers, we have
\begin{equation}
\frac{1}{\sqrt{n}}\sum_{i=1}^{n}|i\rangle_{1}\frac{1}{\sqrt{a}}\sum_{k=0}^{p}\sqrt{a_{k}}|k\rangle_{2}|\|\mathbf{x}_{i}\|^{k}\rangle_{3}|\|\mathbf{x}_{i}\|^{2}\rangle_{4}|0^{\log m}\rangle^{\otimes p}_{5}.
\end{equation}

$\rm{(II.4)}$ Perform $\exp(-\lambda x)$ gate (with details given in Appendix~\ref{Sec:model}) on the fourth register to generate
\begin{equation}
\begin{aligned}
&\frac{1}{\sqrt{n}}\sum_{i=1}^{n}|i\rangle_{1}\frac{1}{\sqrt{a}}\sum_{k=0}^{p}\sqrt{a_{k}}|k\rangle_{2}|\|\mathbf{x}_{i}\|^{k}\rangle_{3}|\exp(-\lambda\|\mathbf{x}_{i}\|^{2})\rangle_{4}\\
&|0^{\log m}\rangle^{\otimes p}_{5}.
\end{aligned}
\end{equation}

$\rm{(II.5)}$ Use $\mathbf{Lemma 3}$ for the third and fourth registers and uncompute redundant registers, we obtain
\begin{equation}
\begin{aligned}
&\frac{1}{\sqrt{n}}\sum_{i=1}^{n}|i\rangle_{1}\frac{1}{\sqrt{a}}\sum_{k=0}^{p}\sqrt{a_{k}}|k\rangle_{2}|\exp(-\lambda\|\mathbf{x}_{i}\|^{2})\|\mathbf{x}_{i}\|^{k}\rangle_{3}\\
&|0^{\log m}\rangle^{\otimes p}_{5}.
\end{aligned}
\end{equation}

$\rm{(II.6)}$ Add one qubit and rotate it from $|0\rangle$ to $[\frac{\exp(-\lambda\|\mathbf{x}_{i}\|^{2})\|\mathbf{x}_{i}\|^{k}}{C}|0\rangle+\sqrt{1-[\frac{\exp(-\lambda\|\mathbf{x}_{i}\|^{2})\|\mathbf{x}_{i}\|^{k}}{C}]^{2}}|1\rangle]_{0}$ controlled on the third register $|\exp(-\lambda\|\mathbf{x}_{i}\|^{2})\|\mathbf{x}_{i}\|^{k}\rangle$, where
$C=\max_{i}[\exp(-\lambda\|\mathbf{x}_{i}\|^{2})\|\mathbf{x}_{i}\|^{k}]$. The above operation is denoted as a controlled rotation operator $R$. Then we uncompute the third register to obtain
\begin{equation}
\begin{aligned}
&\frac{1}{\sqrt{n}}\sum_{i=1}^{n}|i\rangle_{1}\frac{1}{\sqrt{a}}\sum_{k=0}^{p}\sqrt{a_{k}}|k\rangle_{2}[\frac{\exp(-\lambda\|\mathbf{x}_{i}\|^{2})\|\mathbf{x}_{i}\|^{k}}{C}|0\rangle\\
&+\sqrt{1-[\frac{\exp(-\lambda\|\mathbf{x}_{i}\|^{2})\|\mathbf{x}_{i}\|^{k}}{C}]^{2}}|1\rangle]_{0}|0^{\log m}\rangle^{\otimes p}_{5}.
\end{aligned}
\end{equation}

$\rm{(II.7)}$ Apply the quantum amplitude amplification algorithm \cite{BGHP2002} to produce
\begin{equation}
\frac{1}{\sqrt{\Upsilon}}\sum_{i=1}^{n}|i\rangle_{1}\sum_{k=0}^{p}\sqrt{a_{k}}\exp(-\lambda\|\mathbf{x}_{i}\|^{2})\|\mathbf{x}_{i}\|^{k}|k\rangle_{2}|0^{\log m}\rangle^{\otimes p}_{5},
\end{equation}
where $\Upsilon=\sum_{i=1}^{n}\sum_{k=0}^{p}a_{k}[\exp(-\lambda\|\mathbf{x}_{i}\|^{2})]^{2}\|\mathbf{x}_{i}\|^{2k}$.

$\rm{(II.8)}$ Perform the controlled unitary operator $R_{U}=\sum_{k=0}^{p}|k\rangle_{2}\langle k|_{2}\otimes (I^{p-k}U^{k})_{1,5}$, the system becomes
\begin{equation}
\begin{aligned}
&\frac{1}{\sqrt{\Upsilon}}\sum_{i=1}^{n}|i\rangle_{1}\sum_{k=0}^{p}\sqrt{a_{k}}\exp(-\lambda\|\mathbf{x}_{i}\|^{2})\|\mathbf{x}_{i}\|^{k}|k\rangle_{2}\\
&|0^{\log m}\rangle^{\otimes p-k}_{5}|\mathbf{x}_{i}\rangle^{\otimes k}_{5}:=\frac{1}{\sqrt{\Upsilon}}\sum_{i=1}^{n}|i\rangle_{1}|\Psi(\mathbf{x}_{i})\rangle_{2,5}:=|\Psi\rangle,
\end{aligned}
\end{equation}
where
\begin{equation}
\begin{aligned}
&|\Psi(\mathbf{x}_{i})\rangle_{2,5}=\exp(-\lambda\|\mathbf{x}_{i}\|^{2})(\sqrt{a_{0}}|0\rangle|0^{\log m}\rangle^{\otimes p}+\sqrt{a_{1}}|1\rangle\\
&|0^{\log m}\rangle^{\otimes p-1}\|\mathbf{x}_{i}\||\mathbf{x}_{i}\rangle+\cdots+\sqrt{a_{p}}|p\rangle\|\mathbf{x}_{i}\|^{p}|\mathbf{x}_{i}\rangle^{\otimes p})_{2,5}.
\end{aligned}
\end{equation}
Note that
\begin{equation}
\begin{aligned}
&\langle\Psi(\mathbf{x}_{i})|\Psi(\mathbf{x}_{j})\rangle=\exp[-\lambda(\|\mathbf{x}_{i}\|^{2}+\|\mathbf{x}_{j}\|^{2})][a_{0}+a_{1}\\
&\|\mathbf{x}_{i}\|\|\mathbf{x}_{j}\|\langle\mathbf{x}_{i}|\mathbf{x}_{j}\rangle+\cdots+a_{p}(\|\mathbf{x}_{i}\|\|\mathbf{x}_{j}\|\langle\mathbf{x}_{i}|\mathbf{x}_{j}\rangle)^{p}]=w_{ij}.
\end{aligned}
\end{equation}

$\rm{(II.9)}$ Take partial trace for the second and fifth registers, we get
\begin{equation}
\begin{aligned}
\rho_{1}:={\rm Tr}(|\Psi\rangle\langle\Psi|)_{2,5}&=\frac{1}{\Upsilon}\sum_{i=1}^{n}\langle\Psi(\mathbf{x}_{i})|\Psi(\mathbf{x}_{j})\rangle|i\rangle_{1}\langle j|_{1}\\
&=(W+I)/{\rm Tr}(W+I).
\end{aligned}
\end{equation}

\begin{figure}
\centering
	\includegraphics[width=8.5cm]{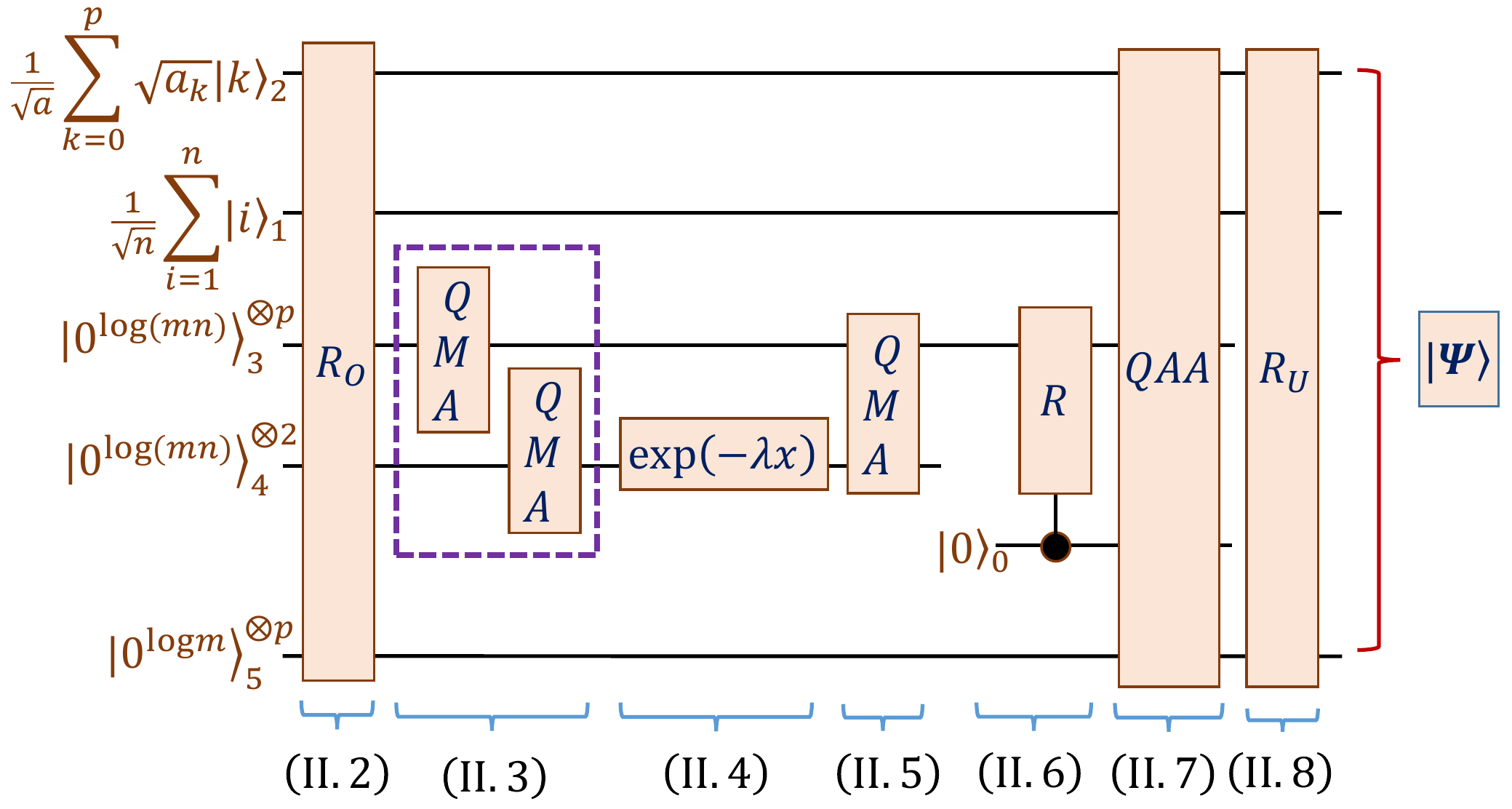}
	\caption{The quantum circuit that generates the quantum state $|\Psi\rangle$. Here the subscripts $0,1,\cdots,5$ represent the index of the registers. $R_{O}$ and $R_{U}$ denote the controlled unitary operator $R_{O}$ and $R_{U}$, respectively. $\rm{QMA}$ denotes the quantum multiply-adder in $\mathbf{Lemma 1}$. $R$ denotes a controlled rotation operator, and $\rm{QAA}$ represents the quantum amplitude amplification algorithm. $\exp(-\lambda x)$ stands for the $\exp(-\lambda x)$ gate. The numbers $\rm{(II.2)}-\rm{(II.9)}$ represent the sequential steps of algorithm.}
	\label{FIG:3}
\end{figure}

According to $\mathbf{Lemma 1}$, the process of generating the state $|\Psi\rangle$ can be regard as an $(a_{1}+s_{1})$-qubits unitary operator $G_{1}$ which implement $G_{1}|0\rangle^{\otimes a_{1}}|0\rangle^{\otimes s_{1}}\mapsto|\Psi\rangle$, s.t.
$\rm {Tr}_{2,5}(|\Psi\rangle\langle\Psi|)=\rho_{1}$, where $a_{1}=\log(p+1)+(p+2)\log(mn)+p\log m+1,s_{1}=\log n$. Then
$V_{1}:=(G_{1}^{\dagger}\otimes I_{1,3,4,0})( I_{2,5}\otimes \rm{SWAP}_{1,3,4,0})(G_{1}\otimes I_{1,3,4,0})$ is a $(1,a_{1}+s_{1},2\epsilon_{1})$-block-encoding of $\rho_{1}$, where $\epsilon_{1}$ is the error that produces the state $|\Psi\rangle$. See $\textbf{(II.2)}$ for detailed analysis for  $\epsilon_{1}$. The whole quantum circuit is shown in Fig.~\ref{FIG:3}.

\textbf{(II.2) The complexty of algorithm in the case of $\|\mathbf{x}_{i}\|\neq1,i=1,\cdots,n$}

The complexity of algorithm is dominated by the controlled unitary operators $R_{O}$ and $R_{U}$, and quantum amplitude amplification algorithm of step $\rm{(II.7)}$. According to the quantum circuits of $R_{O}$ and $R_{U}$, we can obtain that the complexity of $R_{O}$ and $R_{U}$ is $O(p\cdot\poly\log(mn))$ and $O(p\cdot\poly\log(mn/\varepsilon_{x}))$, respectively.

Next we analyze the complexity of step $\rm{(II.7)}$. According to Eq.$(25)$, we can get the probability amplitude of the auxiliary qubit $|0\rangle$ is
\begin{equation}
\frac{\sum_{i=1}^{n}\sum_{k=0}^{p}a_{k}[\exp(-\lambda\|\mathbf{x}_{i}\|^{2})]^{2}\|\mathbf{x}_{i}\|^{2k}}{naC^{2}}=O(\frac{1}{aC^{2}}).
\end{equation}
The equation above is established by the Taylor expansion of order $p$ of $[\exp(\lambda\|\mathbf{x}_{i}\|^{2})]^{2}$, i.e.,
\begin{equation}
[\exp(\lambda\|\mathbf{x}_{i}\|^{2})]^{2}=\exp(2\lambda\|\mathbf{x}_{i}\|^{2})=\sum_{k=0}^{p}a_{k}\|\mathbf{x}_{i}\|^{2k}.
\end{equation}
Thus, we need perform $O(\sqrt{a}C)$ repetitions of quantum amplitude amplification algorithm to obtain the quantum state of Eq.$(26)$.

Besides, we can obtain
\begin{equation}
\sum_{i=1}^{n}\sum_{k=0}^{p}a_{k}[\exp(-\lambda\|\mathbf{x}_{i}\|^{2})]^{2}\|\mathbf{x}_{i}\|^{2k}/n=O(1).
\end{equation}
That means that $\Upsilon=O(n)$.

Finally, we analyze the error that produces the quantum state $|\Psi\rangle$ as follows:
\begin{equation}
\begin{aligned}
&\||\hat{\Psi}\rangle-|\Psi\rangle\|_{2}\\
&=\|\frac{1}{\sqrt{\Upsilon}}\sum_{i=1}^{n}|i\rangle|\hat{\Psi}(\mathbf{x}_{i})\rangle-\frac{1}{\sqrt{\Upsilon}}\sum_{i=1}^{n}|i\rangle|\Psi(\mathbf{x}_{i})\rangle\|_{2}\\
&\leq\frac{1}{\sqrt{\Upsilon}}\sum_{i=1}^{n}\||\hat{\Psi}(\mathbf{x}_{i})\rangle-|\Psi(\mathbf{x}_{i})\rangle\|_{2}\\
&\leq\frac{1}{\sqrt{\Upsilon}}\sum_{i=1}^{n}\sqrt{a}\|\mathbf{x}_{i}\|^{p}[\varepsilon_{a}+(1+2+\cdots+p)\varepsilon_{x}],\\
\end{aligned}
\label{34}
\end{equation}
where $\|\mathbf{x}_{i}\|>1$ and a detailed analysis of the last inequality is
 given in Appendix~\ref{Sec:analysis}. In particular, for $0<\|\mathbf{x}_{i}\|<1$, we also give a detailed analysis in Appendix~\ref{Sec:analysis}.

To ensure that the error of producing $|\Psi\rangle$ is $\epsilon_{1}$, let $\varepsilon_{a}=\varepsilon_{x}$, we get
\begin{equation}
\||\hat{\Psi}\rangle-|\Psi\rangle\|_{2}\leq\sqrt{an}p^{2}(\max_{i}\|\mathbf{x}_{i}\|)^{p}\varepsilon_{x}:=\epsilon_{1}.
\end{equation}
 Thus, we can obtain $1/\varepsilon_{x}=\sqrt{an}p^{2}(\max_{i}\|\mathbf{x}_{i}\|)^{p}/\epsilon_{1}$.

Putting this all together, the complexity is
\begin{equation}
O\{\sqrt{a}pC\poly\log[\sqrt{a}mn^{3/2}p^{2}(\max_{i}\|\mathbf{x}_{i}\|)^{p}/\epsilon_{1}]\}:=O(c_{1}).
\end{equation}

\subsection{Prepare the quantum state to construct the block-encoding of an operator containing the information of $D$}
\label{Sec:D}
The elements of $D$ are $d_{ii}=\sum_{j=1}^{n}w_{ij},i=1,\cdots,n$, that is, the sum of each row of $W$. Thus, $d_{ii}$ can be regarded as the inner product
of the row vector of $W$ and the vector $\mathbf{1}=(1,\cdots,1)^{T}\in R^{n}$. To achieve this, we first give the lemma required.

$\mathbf{Lemma 4}$ \cite{KK2019} (Distance / Inner Products Estimation of two vectors)
Assume that the unitary operators $U|i\rangle|0\rangle=|i\rangle|\mathbf{x}_{i}\rangle$ and $V|j\rangle|0\rangle=|j\rangle|\mathbf{x}_{j}\rangle$
can be performed in time $T$, and the norms $\|\mathbf{x}_{i}\|$ and $\|\mathbf{x}_{j}\|$ are known. Then there is a quantum algorithm can compute
\begin{equation}
|i\rangle|j\rangle|0\rangle\mapsto|i\rangle|j\rangle|\|\mathbf{x}_{i}-\mathbf{x}_{j}\|^{2}\rangle
\end{equation} or
\begin{equation}
\begin{aligned}
|i\rangle|j\rangle&\frac{1}{\sqrt{2}}(|0\rangle|\mathbf{x}_{i}\rangle+|1\rangle|\mathbf{x}_{j}\rangle)|0\rangle\\
&\mapsto|i\rangle|j\rangle\frac{1}{\sqrt{2}}(|0\rangle|\mathbf{x}_{i}\rangle+|1\rangle|\mathbf{x}_{j}\rangle)|\mathbf{x}_{i}\cdot\mathbf{x}_{j}\rangle
\end{aligned}
\end{equation}
with probability at least $1-2\delta$ for any $\delta$ with complexity $O\{[\|\mathbf{x}_{i}\|\|\mathbf{x}_{j}\|T\log(1/\delta)]/\varepsilon\}$, where $\varepsilon$ is the error of $\|\mathbf{x}_{i}-\mathbf{x}_{j}\|^{2}$ or $\mathbf{x}_{i}\cdot\mathbf{x}_{j}$.

$\textbf{1. The process of the quantum algorithm}$

$(1).$ Prepare the quantum state
\begin{equation}
|0\rangle_{1}\frac{1}{n}\sum_{i,j=1}^{n}|i\rangle_{2}|j\rangle_{3}|0^{\log(mn)}\rangle_{4}|0\rangle_{5}|0^{\log(mn)}\rangle_{6}.
\end{equation}

$(2).$ The Distance Estimation algorithm of $\mathbf{Lemma 4}$ is applied to the second, third and fourth registers, we get
\begin{equation}
|0\rangle_{1}\frac{1}{n}\sum_{i,j=1}^{n}|i\rangle_{2}|j\rangle_{3}|\|\mathbf{x}_{i}-\mathbf{x}_{j}\|^{2}\rangle_{4}|0\rangle_{5}|0^{\log(mn)}\rangle_{6}.
\end{equation}

$(3).$ Perform $\exp(-\lambda x)$ gate (See the detailed analysis in Appendix~\ref{Sec:model}) for the fourth register to produce
\begin{equation}
|0\rangle_{1}\frac{1}{n}\sum_{i,j=1}^{n}|i\rangle_{2}|j\rangle_{3}|\exp(-\lambda\|\mathbf{x}_{i}-\mathbf{x}_{j}\|^{2})\rangle_{4}|0\rangle_{5}|0^{\log(mn)}\rangle_{6}.
\end{equation}
In fact, according to Eq.$(1)$, we have $w_{ij}=0$ when $i=j$. However, for Eq.$(41)$, we have $w_{ij}=1,i=j$. Therefore, we need to perform quantum amplitude amplification algorithm to discard $w_{ij}=1,i=j$.

For convenience, we rewrite Eq.$(41)$ as
\begin{equation}
\begin{aligned}
&|0\rangle_{1}[\sqrt{\frac{n^{2}-n}{n^{2}}}\cdot\frac{1}{\sqrt{n^{2}-n}}\sum_{i\neq j,i,j=1}^{n}|i\rangle|j\rangle|w_{ij}\rangle\\
&+\sqrt{\frac{n}{n^{2}}}\cdot\frac{1}{\sqrt{n}}\sum_{i=j=1}^{n}|i\rangle|j\rangle|w_{ij}\rangle]_{2,3,4}|0\rangle_{5}|0^{\log(mn)}\rangle_{6}.
\end{aligned}
\end{equation}

$(4).$ Run the quantum amplitude amplification algorithm to generate
\begin{equation}
|0\rangle_{1}\frac{1}{\sqrt{n^{2}-n}}\sum_{i\neq j,i,j=1}^{n}|i\rangle_{2}|j\rangle_{3}|w_{ij}\rangle_{4}|0\rangle_{5}|0^{\log(mn)}\rangle_{6}.
\end{equation}

$(5).$ Perform Hadamard gate $H$ on the first register
\begin{equation}
\begin{aligned}
&\frac{1}{\sqrt{2}}(|0\rangle+|1\rangle)_{1}\frac{1}{\sqrt{n^{2}-n}}\sum_{i\neq j,i,j=1}^{n}|i\rangle_{2}|j\rangle_{3}|w_{ij}\rangle_{4}|0\rangle_{5}\\
&|0^{\log(mn)}\rangle_{6}.
\end{aligned}
\end{equation}

$(6).$ Apply the controlled unitary operator $|0\rangle\langle0|_{1}\otimes I _{2,3}\otimes R_{w(4,5)}\otimes I_{6}+|1\rangle\langle1|_{1}\otimes I_{2-6}$, where $R_{w}$ is a controlled rotation operator, and uncompute the fourth register, we get
\begin{equation}
\begin{aligned}
&\frac{1}{\sqrt{n}}\sum_{i=1}^{n}|i\rangle_{2}\frac{1}{\sqrt{2}}[|0\rangle_{1}\frac{1}{\sqrt{n-1}}\sum_{j\neq i,j=1}^{n}|j\rangle_{3}(w_{ij}|0\rangle+\\
&\sqrt{1-w_{ij}^{2}}|1\rangle)_{5}+|1\rangle_{1}\frac{1}{\sqrt{n-1}}\sum_{j\neq i,j=1}^{n}|j\rangle_{3}|0\rangle_{5}]|0^{\log(mn)}\rangle_{6}\\
&:=\frac{1}{\sqrt{n}}\sum_{i=1}^{n}|i\rangle_{2}\frac{1}{\sqrt{2}}(|0\rangle_{1}|\varphi\rangle_{3,5}+|1\rangle_{1}|\psi\rangle_{3,5})|0^{\log(mn)}\rangle_{6},\\
\end{aligned}
\end{equation}
where
\begin{equation}
\begin{aligned}
&|\varphi\rangle_{3,5}=\frac{1}{\sqrt{n-1}}\sum_{j\neq i,j=1}^{n}|j\rangle_{3}(w_{ij}|0\rangle+\sqrt{1-w_{ij}^{2}}|1\rangle)_{5},\\
&|\psi\rangle_{3,5}=\frac{1}{\sqrt{n-1}}\sum_{j\neq i,j=1}^{n}|j\rangle_{3}|0\rangle_{5}.
\end{aligned}
\end{equation}

Note that the inner products of $|\varphi\rangle$ and $|\psi\rangle$ is
\begin{equation}
\langle\varphi|\psi\rangle=\frac{\sum_{j\neq i,j=1}^{n}w_{i,j}}{n-1}=\frac{d_{ii}}{n-1}.
\end{equation}

$(7).$ Apply the Inner Products Estimation algorithm
 of $\mathbf{Lemma 4}$, we have
\begin{equation}
\frac{1}{\sqrt{n}}\sum_{i=1}^{n}|i\rangle_{2}\frac{1}{\sqrt{2}}(|0\rangle_{1}|\varphi\rangle_{3,5}+|1\rangle_{1}|\psi\rangle_{3,5})|\langle\varphi|\psi\rangle\rangle_{6}.
\end{equation}

$(8).$ Attach a register, then perform a controlled rotation operator $R_{p}$ and uncompute the first, third, fifth and sixth registers to produce
\begin{equation}
\begin{aligned}
\frac{1}{\sqrt{n}}\sum_{i=1}^{n}|i\rangle_{2}[&\sqrt{\langle\varphi|\psi\rangle}|00^{\log (n-1)}\rangle\\
&+\sqrt{1-\langle\varphi|\psi\rangle}|10^{\log(n-1)}\rangle]_{0}.
\end{aligned}
\end{equation}

$(9).$ Run the quantum amplitude amplification algorithm
to get
\begin{equation}
\frac{1}{\sqrt{\tau}}\sum_{i=1}^{n}\sqrt{d_{ii}}|i\rangle_{2}|0^{\log n}\rangle_{0},
\end{equation}
where $\tau=\sum_{j\neq i,i,j=1}^{n}\exp(-\lambda\|\mathbf{x}_{i}-\mathbf{x}_{j}\|^{2})$.

$(10).$ Apply CNOT gate to the zeroth register
\begin{equation}
\frac{1}{\sqrt{\tau}}\sum_{i=1}^{n}\sqrt{d_{ii}}|i\rangle_{2}|i\rangle_{0}:=|\phi\rangle.
\end{equation}

$(11).$ Take partial trace for the zeroth register to get
\begin{equation}
\rho_{2}:={\rm Tr}_{0}(|\phi\rangle\langle\phi|)=\frac{1}{\tau}\sum_{i=1}^{n}d_{ii}|i\rangle_{2}\langle i|_{2}=\frac{D}{{\rm Tr}(D)}.
\end{equation}

\begin{figure}
\centering
	\includegraphics[width=8.5cm]{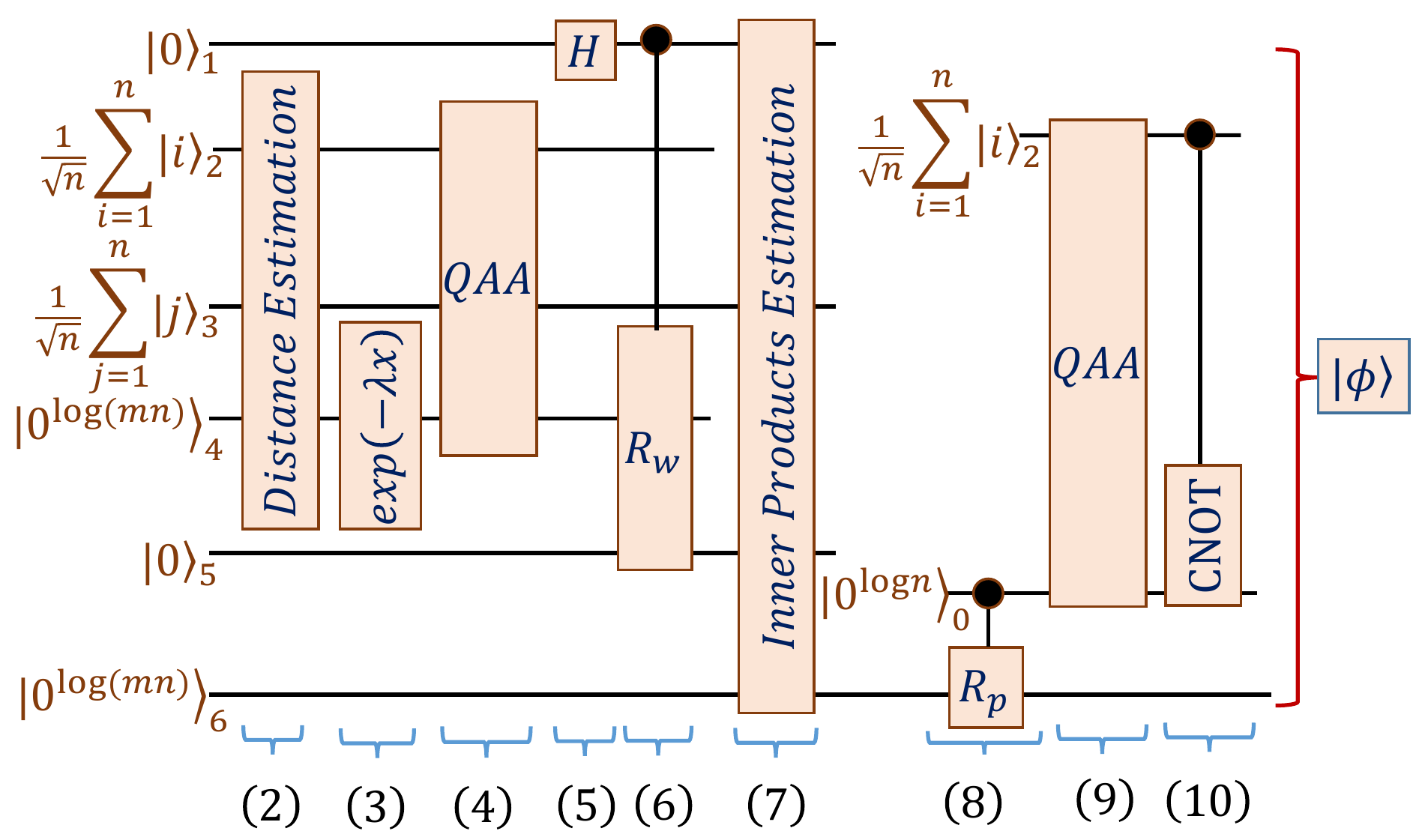}
	\caption{The quantum circuit that produces the quantum state $|\phi\rangle$. Here the subscripts $0,1,\cdots,6$ represent the index of the registers. Distance Estimation and Inner Products Estimation represent the algorithms in $\mathbf{Lemma 2}$, $\exp(-\lambda x)$ stands for the $\exp(-\lambda x)$ gate. $\rm{QAA}$ represents the quantum amplitude amplification algorithm, $H$ denotes the Hadamard operation, $R_{w}$ and $R_{p}$ denote the controlled rotation operator, and $\rm{CNOT}$ denotes the $\rm{CNOT}$ quantum gate. The numbers $\rm{(2)}-\rm{(10)}$ represent the sequential steps of algorithm.}
	\label{FIG:4}
\end{figure}

According to $\mathbf{Lemma 1}$, the process of producing the quantum state $|\phi\rangle$ also can be regarded as an $(a_{2}+s_{2})$-qubits unitary operator $G_{2}$ which realize $G_{2}|0\rangle^{\otimes a_{2}}|0\rangle^{\otimes s_{2}}\mapsto|\phi\rangle$, s.t.,
$\rm {Tr}_{0}(|\phi\rangle\langle\phi|)=\rho_{2}$, where $a_{2}=2(1+\log n+\log(mn))$, $s_{2}=\log n$. Thus
$V_{2}:=(G_{2}^{\dagger}\otimes I_{1-6})( I_{0}\otimes \rm{SWAP}_{1-6})(G_{2}\otimes I_{1-6})$ is an
$(1,a_{2}+s_{2},2\epsilon_{2})$-block-encoding of $\rho_{2}$, where $\epsilon_{2}$ is the error that generates the state $|\phi\rangle$, and $\epsilon_{2}$ is analyzed in detail in the complexity analysis of the algorithm. The whole quantum circuit is shown in Fig.~\ref{FIG:4}.

$\textbf{2. The complexity of the algorithm}$

In step$(1)$, we perform $2\log n$ Hadamard gates on the second and third registers of the initial state
$|0\rangle_{1}|0^{\log n}\rangle_{2}|0^{\log n}\rangle_{3}|0^{\log(mn)}\rangle_{4}|0\rangle_{5}|0^{\log(mn)}\rangle_{6}$ to produce the quantum state of Eq.$(39)$. Thus, the complexity of step$(1)$ is $2\log n$.

In step$(2)$, according to $\mathbf{Lemma 4}$, the complexity is
\begin{equation} O\{[(\max_{i}\|\mathbf{x}_{i}\|)^{2}\poly\log(mn/\epsilon_{x})\log(1/\delta_{1})]/\varepsilon_{d}\},
\end{equation}
where $1-2\delta_{1}$ is the probability of success of the algorithm with any $\delta_{1}$ and $\varepsilon_{d}$ is the error of $\|\mathbf{x}_{i}-\mathbf{x}_{j}\|^{2}$.

In step$(3)$, the complexity of the $\exp(-\lambda x)$ gate is $O[\poly\log(1/\varepsilon_{d})]$ (with details given in Appendix~\ref{Sec:model}) which is smaller than $O(1/\varepsilon_{d})$ caused by the step$(2)$. Therefore we can ignore the complexity of these gates.

In step$(4)$, the probability amplitude of the target states of Eq.$(42)$ is $p=(n^{2}-n)/n^{2}=O(1)$. Thus, we perform $O(1)$ times of quantum amplitude amplification algorithm to produce the state in Eq.$(43)$.

In step$(5)$-step$(6)$, it contains one Hadamard gate and a controlled rotation operator $R_{w}$ which has complexity $O(1)$.

In step$(7)$, we analyze the error in calculating $\langle\varphi|\psi\rangle$ is
\begin{equation}
\begin{aligned}
|\hat{\langle\varphi|\psi\rangle}-\langle\varphi|\psi\rangle|
&\leq\frac{1}{n-1}\sum_{j\neq i,j=1}^{n}|\exp(-\lambda\hat{\omega})-\exp(-\lambda\omega)|\\
&\leq\frac{1}{n-1}\sum_{j\neq i,j=1}^{n}|\lambda\exp(-\lambda\xi)(\hat{\omega}-\omega)|\\
&\leq\frac{1}{n-1}\sum_{j\neq i,j=1}^{n}\lambda|\hat{\omega}-\omega|\leq\lambda\varepsilon_{d},
\end{aligned}
\end{equation}
where $\omega=\|\mathbf{x}_{i}-\mathbf{x}_{j}\|^{2}$, the second inequality comes from the Lagrange's mean value theorem \cite{PKT1998}, $\xi$ takes value from $\hat{\omega}$
to $\omega$.

According to $\mathbf{Lemma 4}$, we can get that the complexity is
\begin{equation} O\{[(\max_{i}\|\mathbf{x}_{i}\|)^{2}\poly\log(mn/\varepsilon_{x})\upsilon]/(\lambda\varepsilon_{d}^{2})\},
\end{equation}
where $\upsilon=\log(1/\delta_{1})\log(1/\delta_{2})$,
$1-2\delta_{2}$ is the probability of success of the algorithm with any $\delta_{2}$ and $\lambda\varepsilon_{d}$ is the error of $\langle\varphi|\psi\rangle$.

In step$(8)$-step$(10)$, the complexity comes mainly from quantum amplitude amplification algorithm. The probability amplitude of the target state in Eq.(49) is
\begin{equation}
\begin{aligned}
p_{0}&=[\sum_{j\neq i,i,j=1}^{n}\exp(-\lambda\|\mathbf{x}_{i}-\mathbf{x}_{j}\|^{2})]/[n(n-1)]\\
&\geq\min_{i,j}[\exp(-\lambda\|\mathbf{x}_{i}-\mathbf{x}_{j}\|^{2})]:=r.
\end{aligned}
\end{equation}
Thus, we need $O(1/\sqrt{r})$ applications of quantum amplitude amplification algorithm to generate the quantum state in Eq.$(50)$. In addition, we can get $p_{0}={\rm Tr}(D)/[n(n-1)]$, i.e., ${\rm Tr}(D)=n(n-1)p_{0}$.

Finally, we analyze the error that produces the quantum state $|\phi\rangle$ is
\begin{equation}
\begin{aligned}
\||\hat{\phi}\rangle-|\phi\rangle\|_{2}^{2}&=\|\frac{1}{\sqrt{\tau}}\sum_{i=1}^{n}[\sqrt{\hat{d_{ii}}}-\sqrt{d_{ii}}]|i\rangle_{2}|i\rangle_{0}\|_{2}^{2}\\
&=\frac{1}{\tau}\sum_{i=1}^{n}(n-1)[\sqrt{\hat{\langle\varphi|\psi\rangle}}-\sqrt{\langle\varphi|\psi\rangle}]^{2}\\
&=\frac{1}{\tau}\sum_{i=1}^{n}(n-1)[\frac{1}{2\sqrt{\xi_{0}}}(\hat{\langle\varphi|\psi\rangle}-\langle\varphi|\psi\rangle)]^{2}\\
&\leq\frac{1}{\tau}\sum_{i=1}^{n}(n-1)\frac{1}{4\xi_{0}}\lambda^{2}\varepsilon_{d}^{2}\leq\lambda^{2}\varepsilon_{d}^{2}/4r,
\end{aligned}
\end{equation}
where the second equation comes from Eq.$(47)$, the third equation follows from the Lagrange's mean value theorem \cite{PKT1998}, $\xi_{0}$ takes value from $\hat{\langle\varphi|\psi\rangle}$
to $\langle\varphi|\psi\rangle$, and the last inequality holds by
$\tau\geq n(n-1)r$.

The complexity of each step of the algorithm is shown in Table.~\ref{Table}.

\begin{table}[htbp]
\centering
\label{TAB:table}
\caption{The complexity of each step of the algorithm }
\begin{tabular}{cccccccc|cc}
\hline steps & complexity \\
\hline  $(1)$ & $2\log n$ \\
$(2)$ & $O\{[(\max_{i}\|\mathbf{x}_{i}\|)^{2}\poly\log(mn/\varepsilon_{x})\log(1/\delta_{1})]/\varepsilon_{d}\}$ \\
$(3)$ & $O(\poly\log(1/\varepsilon_{d}))$  \\
$(4)-(6)$ & $O(1)$  \\
$(7)$ & $O\{[(\max_{i}\|\mathbf{x}_{i}\|)^{2}\poly\log(mn/\varepsilon_{x})\upsilon]/(\lambda\varepsilon_{d}^{2})\}$ \\
$(8)-(10)$ & $O(1/\sqrt{r})$ \\
 \hline\hline
\end{tabular}
\label{Table}
\end{table}

To make the error of $|\phi\rangle$ equal to $\epsilon_{2}$, we have
\begin{equation}
\||\hat{\phi}\rangle-|\phi\rangle\|_{2}\leq\lambda\varepsilon_{d}/2\sqrt{r}:=\epsilon_{2}.
\end{equation}
Thus we can obtain $1/\varepsilon_{d}=\lambda/2\sqrt{r}\epsilon_{2}$.

Putting all complexity together and letting $\varepsilon_{x}=\varepsilon_{d}$, $\delta_{1}=\delta_{2}=O(\poly\log n)$, we get
 \begin{equation}
 O[\frac{\lambda(\max_{i}\|\mathbf{x}_{i}\|)^{2}\poly\log[(\lambda mn)/(\sqrt{r}\epsilon_{2})]}{4r^{3/2}\epsilon_{2}^{2}}]:=O(c_{2}).
\end{equation}

\subsection{Implement the quantum simulation of $L$}
\label{Sec:I}

Here, we first implement the quantum simulation of $L$ in the case of $\|\mathbf{x}_{i}\|\neq1,i=1,\cdots,n$. Similarly, we provide a detailed analysis for the case of $\|\mathbf{x}_{i}\|=1$ in Appendix~\ref{Sec:CI}.

According to $L=D-W$, we have
\begin{equation}
\mathcal{L}=\frac{L}{{\rm Tr}(L)}=\frac{D}{{\rm Tr}(D)}-\frac{W}{{\rm Tr}(D)},
\end{equation}
where the last equation comes from ${\rm Tr}(L)={\rm Tr}(D)$.

Due to $\rho_{1}=(W+I)/{\rm Tr}(W+I)$ and $\rho_{2}=D/{\rm Tr}(D)$, we can obtain
\begin{equation}
\begin{aligned}
\mathcal{L}&=\rho_{2}-\frac{{\rm Tr}(W+I)}{{\rm Tr}(D)}\rho_{1}+\frac{I}{{\rm Tr}(I)}\frac{{\rm Tr}(I)}{{\rm Tr}(D)}.
\end{aligned}
\end{equation}

For $I/{\rm Tr}(I)$, we can prepare the quantum state $|\tau\rangle=\frac{1}{\sqrt{n}}\sum_{i=1}^{n}|i\rangle_{1}|i\rangle_{2}$ can be prepared with $O(\log n)$ Hardmard and CNOT gates. Thus, we have
\begin{equation}
\begin{aligned}
\frac{I}{{\rm Tr}(I)}=
{\rm Tr}_{2}(|\tau\rangle\langle\tau|)=\frac{1}{n}\sum_{i=1}^{n}|i\rangle_{1}\langle i|_{1}:=\rho_{3}.
\end{aligned}
\end{equation}

According to $\mathbf{Lemma 1}$, we also get an $(1,2\log n,0)$-block-encoding of $\rho_{3}$, that is,
$V_{3}:=(G_{3}^{\dagger}\otimes I_{1})(I_{2}\otimes \rm{SWAP}_{1})(G_{3}\otimes I_{1})$ where
$G_{3}|0^{\log n}\rangle|0^{\log n}\rangle\mapsto|\tau\rangle$, s.t.,
$\rm {Tr}_{2}(|\tau\rangle\langle\tau|)=\rho_{3}$.

Thus, we obtain
\begin{equation}
\mathcal{L}=-c\rho_{1}+\rho_{2}+c\rho_{3},
\end{equation}
where $c={\rm Tr}(W+I)/{\rm Tr}(D)={\rm Tr}(I)/{\rm Tr}(D),0<c<1$, and ${\rm Tr}(D)$ can be effectively evaluated in Sec.~\ref{Sec:D}. This can be viewed as a linear combination of block-encoded operators.

 By $\mathbf{Definition 2}$, let $\mathbf{y}=(-c,1,c)$ and $\|\mathbf{y}\|_{1}\leq\beta=3$. Let
 $b=2, c_{j}=d_{j}=\sqrt{y_{j}}$, $j=1,2,3$. We can effectively construct an $(3,2,\varepsilon_{y})$-state-preparation-pair $(P_{L},P_{R})$ of $\mathbf{y}$ that satisfies the requirements of $\mathbf{Definition 2}$ \cite{G2002}.

In addition, we construct a $(\log n+l+2)$-qubits unitary
$Q=\sum_{j=1}^{3}|j\rangle\langle j|\otimes V_{j}+(I-\sum_{j=1}^{3}|j\rangle\langle j|)\otimes I_{l}\otimes I_{\log n}$ such that for $j=1,2,3$, where $l=O(p\log(pmn))$ comes from $\max\{a_{1}+s_{1},a_{2}+s_{2}\}$, we have that $V_{j}$ is an $(1,l,\varepsilon_{l})$-block-encoding of $\rho_{j}$, where $\varepsilon_{l}=\min\{\epsilon_{1},\epsilon_{2}\}$.

 According to $\mathbf{Lemma2}$, we can implement unitary $G$ which is an $(3,l+2,\varepsilon_{y}+3\varepsilon_{l})$-block-encoding of $\mathcal{L}$, with a single use of $Q$, $P_{L}$, and $P_{R}$.

Combining $\mathbf{Theorem 1}$, we can implement an $\varepsilon$-precise the Hamiltonian simulation unitary operator which is an $(1,l+4,\varepsilon)$-block-encoding of $\exp(-i\mathcal{L}t)$ with
\begin{equation}
O[3t+\log(1/\varepsilon)/\log\log(1/\varepsilon)]
\end{equation}
uses of controlled-$G$ or its inverse and with $O[3(l+2)t+(l+2)\log(1/\varepsilon)/\log\log(1/\varepsilon)]$ two-qubit gates, where $\varepsilon=2t(\varepsilon_{y}+3\varepsilon_{l})$.

\subsection{Extract the eigeninformation of $L$}
\label{Sec:Extract}

In this section, we use quantum phase estimation algorithm to extract the
$1\leq d\leq(n-1)$ non-zero eigenvalues and eigenvectors of $\mathcal{L}$.

Suppose that the eigendecomposition form of $\mathcal{L}$ is
\begin{equation}
 \mathcal{L}=\sum_{j=1}^{n}\gamma_{j}|\mathbf{u}_{j}\rangle\langle\mathbf{u}_{j}|,
\end{equation}
where $\{\gamma_{j}\}_{j=1}^{n}$ and $\{|\mathbf{u}_{j}\rangle\}_{j=1}^{n}$ are the eigenvalues and the corresponding eigenvectors of $\mathcal{L}$, respectively.

Our algorithm works as the following steps.

$\textbf{(1) The steps of the quantum algorithm}$

$1$. Perform Hadamard and $\rm{CNOT}$ gates on the initial states $|0^{\log n}\rangle_{1}|0^{\log n}\rangle_{2}|0^{\log n}\rangle_{3}$ to produce the quantum state
\begin{equation}
|\omega\rangle=\frac{1}{\sqrt{n}}\sum_{j=1}^{n}|j\rangle_{1}|j\rangle_{2}\otimes|0^{\log n}\rangle_{3}.
\end{equation}

$2$. Take partial trace for the second register to get
\begin{equation}
{\rm{Tr}}_{2}(|\omega\rangle\langle\omega|)=\frac{1}{n}\sum_{j=1}^{n}|j\rangle_{1}\langle j|_{1}\otimes|0^{\log n}\rangle_{3}\langle0^{\log n}|_{3}.
\end{equation}
Although we don't know the information of the state $|\mathbf{u}_{j}\rangle,j=1,\cdots,n$ in advance, we have
\begin{equation}
\frac{1}{n}\sum_{j=1}^{n}|j\rangle\langle j|=\frac{I}{{\rm Tr}(I)}=\frac{1}{n}\sum_{j=1}^{n}|\mathbf{u}_{j}\rangle\langle\mathbf{u}_{j}|.
\end{equation}
Therefore, we can obtain
\begin{equation}
\begin{aligned}
&\frac{1}{n}\sum_{j=1}^{n}|j\rangle\langle j|\otimes|0^{\log n}\rangle\langle0^{\log n}|\\
&=\frac{1}{n}\sum_{j=1}^{n}|\mathbf{u}_{j}\rangle\langle\mathbf{u}_{j}|\otimes|0^{\log n}\rangle\langle0^{\log n}|.
\end{aligned}
\end{equation}

$3$. Run the quantum phase estimation algorithm by simulating $\exp(-i\mathcal{L}t)$ to reveal the eigenvalues and eigenvectors of $\mathcal{L}$
\begin{equation}
\frac{1}{n}\sum_{j=1}^{n}|\mathbf{u}_{j}\rangle\langle\mathbf{u}_{j}|\otimes|\gamma_{j}\rangle\langle\gamma_{j}|.
\end{equation}

$4$. Use the quantum algorithm for finding the minimum  to reveal the $d$ minimized nonzero eigenvalues $\gamma_{j}$, and the corresponding eigenvectors $|\mathbf{u}_{j}\rangle$, $j=1,\cdots,d$ \cite{D1996,P2022}.

$\textbf{(2) The complexity of the algorithm}$

In step $1$, the complexity is $O(\poly\log n)$, which  comes from Hadamard and $\rm{CNOT}$ gates.

In step $3$, according to Ref.\cite{HHL2009}, it takes $t=O(1/\epsilon)$ times to yields the eigeninformation of $\mathcal{L}$ with accuracy $\epsilon$. Therefore, the complexity of the algorithm is
\begin{equation}
O(\max\{c_{1},c_{2}\}/3\epsilon),
\end{equation}
where $c_{1}$ and $c_{2}$ are shown in Eq.$(36)$ and Eq.$(59)$, which represent the complexity of generating the block-encodings of $W$ and $D$, respectively.

In step $4$, to reveal the $1\leq d\leq(n-1)$ minimized nonzero eigenvalues and the corresponding eigenvectors of $\mathcal{L}$, we need to run $O(d)$ times of the algorithm for find the minimum that output the minimum values with probability larger than $1/2$ with the query complexity $O(\sqrt{n})$ \cite{D1996,P2022}. Thus the total complexity is $O(d\sqrt{n})$.

Putting all the complexity together, we can obtain the complexity of the whole quantum algorithm is
\begin{equation}
O(d\sqrt{n}\max\{c_{1},c_{2}\}/3\epsilon).
\end{equation}

When $a$, $p$, $C$, $\lambda$, $\sqrt{r}$ and $\max_{i}\|\mathbf{x}_{i}\|$ are all $O(1)$, and letting $1/\epsilon_{1}=1/\epsilon_{2}=1/\epsilon=O(\poly\log n)$, our quantum algorithm takes time
\begin{equation}
O[d\sqrt{n}\poly\log(mn^{3/2})].
\end{equation}
It is shown that our algorithm achieve a polynomial speedup on $n$ and an exponential speedup on $m$ compared with the classical algorithm whose complexity is $O(mn^{2}+dn^{3})$.

In addition, our algorithm can also extract the eigeninformation of $W$, which is of great significant \cite{CWWLF2008,GLZA2021,LYWHG2020,ZTLZC2020}. See the detailed analysis in Appendix~\ref{Sec:Re}.

\section{Generalization: Solve the eigenproblem of $L_{s}$ and $L_{r}$}
\label{sec:generalization}

In this section, we extend our quantum algorithm to solve the eigenproblem of $L_{s}$ and $L_{r}$ in Eq.$(3)$. We assume that the eigendecomposition form of $L_{s}$ is
\begin{equation}
L_{s}=\sum_{i=1}^{n}\mu_{i}|\mathbf{v}_{i}\rangle\langle\mathbf{v}_{i}|,
\end{equation}
where $\{\mu_{i}\}_{i=1}^{n}$ and $\{|\mathbf{v}_{i}\rangle\}_{i=1}^{n}$ are the eigenvalues and the corresponding eigenvectors of $L_{s}$, respectively.

According to the properties of $L_{s}$ and $L_{r}$ \cite{CG1997}, we know that the eigenvalues of  $L_{r}$ are also $\{\mu_{i}\}_{i=1}^{n}$ and the corresponding eigenvectors are $\{D^{-\frac{1}{2}}|\mathbf{v}_{i}\rangle\}_{i=1}^{n}$. In particular, $\mu_{1}=0$ is the unique zero eigenvalue of $L_{s}$, and the corresponding eigenvector is  $|\mathbf{v}_{1}\rangle=D^{\frac{1}{2}}\mathbf{1}$. Therefore, when we obtain the eigeninformation of $L_{s}$, we can also get the eigeninformation of $L_{r}$. Next we show that how to extract the eigeninformation of $L_{s}$.

To achieve this, our core task is to first realize the quantum simulation of $L_{s}$. According to Eq.$(3)$, we have
\begin{equation}
L_{s}=[\frac{D}{{\rm Tr}(D)}]^{-\frac{1}{2}}[\frac{L}{{\rm Tr}(L)}][\frac{D}{{\rm Tr}(D)}]^{-\frac{1}{2}}=(\rho_{2})^{-\frac{1}{2}}\mathcal{L}(\rho_{2})^{-\frac{1}{2}},
\end{equation}
where the first equation comes from ${\rm Tr}(L)={\rm Tr}(D)$.

We have constructed the block-encodings of $\rho_{2}$ and $\mathcal{L}$, respectively. Based on $\mathbf{Lemma 4}$ and $\mathbf{Lemma 8}$ of Ref.\cite{CSGJ2018}, we can design the block-encoding of $L_{s}$. For ease of understanding , we write it as  the following lemma, show below:

$\mathbf{Lemma 5}$ (Block-encoding of $A^{-c}BA^{-c}$)

Let $c\in(0,\infty)$, $\varsigma_{1}\in(0,1/2]$, and let $A$ and $B$ are Hermitian matrices, and $A$ satisfy $I/\kappa\preceq A \preceq I$ where $\kappa\geq 2$. Let $\zeta_{1}=O[\frac{\varsigma_{1}}{\kappa^{1+c}\max\{1,c\}\log(\kappa^{c}/\varsigma_{1})\log^{2}(\kappa(c+1)\log(1/\varsigma_{1}))}]$. $U$ is an $(\alpha_{1},b_{1},\zeta_{1})$-block-encoding of $A$ that can be implemented using $T_{U}$ elementary gates and $V$ is an  $(\alpha_{2},b_{2},\zeta_{2})$-block-encoding of $B$ that can be implemented using $T_{V}$ elementary gates. Then we can implement a unitary $F$ that is an $(4\kappa^{2c}\alpha_{2},2b_{1}+b_{2}+O(\log(\kappa^{c}\max(1,c)\log(\kappa^{c}/\varsigma_{1}))),4\kappa^{c}\alpha_{2}\varsigma_{1}+4\kappa^{2c}\zeta_{2})$-block-encoding of $A^{-c}BA^{-c}$ in cost
\begin{equation}
\begin{aligned}
O[&\max\{1,c\}[2\alpha_{1}\kappa\log(\frac{\kappa^{c}}{\varsigma_{1}})(a_{1}+T_{U})\\
&+\kappa\log^{2}((\max\{1,c\}\kappa^{\max\{1,c\}})/\varsigma_{1})]+T_{V}].
\end{aligned}
\end{equation}

 According to $\mathbf{Lemma5}$, for $L_{s}$, we get $c=1/2$, $A=\rho_{2}$, $B=\mathcal{L}$. And we assume that $\rho_{2}$ and $\mathcal{L}$ satisfy the conditions of $\mathbf{Lemma5}$. We have obtained that the unitary operator $V_{2}$ which is an $(1,a_{2}+s_{2},2\epsilon_{2})$-block-encoding of $\rho_{2}$ and the unitary operator $G$ which is an $(3,l+2,\varepsilon_{y}+3\epsilon_{l})$-block-encoding of $\mathcal{L}$, respectively.

 Therefore, we can implement a unitary $F$ that is an
$(12\kappa,2(a_{2}+s_{2}+1)+l+O(\log(\kappa^{\frac{1}{2}}\log(\frac{\kappa^{\frac{1}{2}}}{\varsigma_{1}}))),\epsilon_{f})$-block-encoding of $L_{s}$
 in cost
 \begin{equation}
O[2\kappa\log(\frac{\kappa^{\frac{1}{2}}}{\varsigma_{1}})(a_{2}+s_{2}+c_{2})+\kappa\log^{2}(\frac{\kappa}{\varsigma_{1}})+T_{\mathcal{L}}],
 \end{equation}
where $\epsilon_{f}=12\kappa^{\frac{1}{2}}\varsigma_{1}+4\kappa(\varepsilon_{y}+3\epsilon_{l})$ and $T_{\mathcal{L}}=\max\{c_{1},c_{2}\}$ is the complexity of produces the unitary operator $G$.

Finally, combining $\mathbf{Theorem1}$ and Sec.~\ref{Sec:Extract}, we can obtain the the $d$ minimized nonzero eigenvalues of  $L_{s}$ and the corresponding eigenvectors $\{|\mathbf{v}_{i}\rangle\}_{i=1}^{d}$. To obtain the eigeninformation of $L_{r}$, we perform the quantum technique of $\mathbf{Lemma 30}$ in Ref.\cite{CSGJ2018} to get the state $\frac{(\rho_{2})^{-\frac{1}{2}}|\mathbf{v}_{i}\rangle}{\|(\rho_{2})^{-\frac{1}{2}}|\mathbf{v}_{i}\rangle\|}$ for each $|\mathbf{v}_{i}\rangle$, $i=1,\cdots,d$, that is, the corresponding eigenvectors $\{D^{-\frac{1}{2}}|\mathbf{v}_{i}\rangle\}_{i=1}^{n}$ of $L_{r}$.

\section{Discussion}
\label{Sec:Discussion}

  In Refs.\cite{KIL2021,HYX2016}, scholars use the definition of  $L=BB^{T}$ to design quantum algorithms to solve related problems, and has a significant speedup compared with its classical counterparts, where $B$ is an incidence matrix to store the relationship between each node and its connected edges. As the introduction pointed out, their algorithm cannot efficiently solve the eigenproblems of $L$ of a fully connected weighted graph. A straightforward idea is to design quantum algorithm to efficiently realize the strong assumption of Ref.\cite{HYX2016}, then obtain the eigeninformation of $L_{s}$ by using the existing technique of Ref.\cite{HYX2016}. However, the strong above assumption also require access to the norms of the column vector of  $B$ and $D$, and an efficient quantum algorithm has not yet been found to implement it. In addition, the algorithm of Ref.\cite{HYX2016} uses the Hermitian chain product technique \cite{IL2016} to realize the quantum simulation of $L_{s}$. This makes the complexity of the algorithm have a cubic dependent on the inverse of the simulation error.

 In our algorithm, we adopt the definition of $L=D-W$ to design the quantum algorithm for the following three main reasons: $(1)$ the strong assumption of Ref.\cite{HYX2016} is avoided; $(2)$ the eigeninformation of $W$ is of great significant \cite{CWWLF2008,GLZA2021,LYWHG2020,ZTLZC2020}; $(3)$ solving the eigenproblem of $L_{s}$ and $L_{r}$ also requires access to $D$. In addition, we adopt the optimal Hamiltonian simulation technique based on the block-encoding framework \cite{LGCI2019,GSYLW2019,CSGJ2018} to implement the quantum simulation of $L$, which reduce the algorithm's dependence on simulation error.

 In particular, our algorithm also solve the eigenproblem of $W$ in Appendix~\ref{Sec:Re}. We find that the Gaussian kernel matrix $K$ satisfies $K=W+I$, thus our algorithm can also be used to solve the eigenproblem of $K$. Clearly, our algorithm is a quantum algorithm under the circuit model, while the quantum algorithms proposed in Refs.\cite{CGg2016,S2019,SADNP2021} are formulated with the generalized coherent states, which is the specialized language of quantum optics, to construct $K$. This makes them likely not universal quantum computing paradigm \cite{LZXW2020}. Compared to Ref.\cite{LZXW2020}, our algorithm can not only process the scenario where the modulus length of each sample data point $\mathbf{x}_{i}$ is not equal to $1$, i.e., $\|\mathbf{x}_{i}\|\neq 1$, $i=1,2,\cdots,n$, but also provide the optimal Hamiltonian simulation algorithm based on the block-encoding framework for $K$ to reduce the algorithm's dependence on simulation error. Similarly, our algorithm can also be extended to solve arbitrary nonlinear kernel matrix, which has a wide range of applications in classification, dimensionality reduction, regression and so on. In addition, Sornsaeng et al.\cite{SADNP2021} proposed using the quantum matrix algebra toolbox \cite{zhao2021} algorithm to quantum simulation of $D$, and achieved a certain acceleration effect. Compared with their algorithm, our algorithm has significant improvement in both simulation time and simulation error.

\section{Conclusion}
\label{Sec:Conclusion}

In summary, we designed an efficient quantum algorithm to solve the eigenproblem of $L$ of a fully connected weighted graph. Specifically, we designed special controlled unitary operators to construct the block-encodings of operators containing the information of $W$ and $D$ respectively, and further obtain the block-encoding of $L$. Then we employed the optimal Hamiltonian simulation technology based on the block-encoding framework to realize the quantum simulation of $L$. Finally, we adopted the quantum phase estimation algorithm to extract the eigenvalues and eigenvectors of $L$. It is shown that compared with the classical algorithm, our quantum algorithm achieve polynomial speedup in the number of vertices and exponential speedup in the dimension of each vertex. Additionally, we also extended our algorithm to solve the eigenproblem of $W$, $L_{s}$ and $L_{r}$.

We expect that our quantum algorithm and the techniques mentioned, such as the quantum  technique for constructing the block-encoding of an operator and the analysis of the error propagation of the quantum state, can provide new ideas for quantum algorithms to solve other problems. Furthermore, exploring the application of our quantum algorithms to real data is a goal worth considering in the future.

Besides, the advantages of our algorithms usually rely on a fault-tolerant quantum computer, which may take a long time horizon to implement. Recently, a few scholars have employed variational quantum algorithms \cite{PMS2014,CMAA2021} which can be implemented on the Noisy Intermediate-Scale Quantum (NISQ) devices \cite{P2018} to solve several problems related to $L$. In $2020$, Slimane et al. proposed a variational Laplacian eigenmap algorithm \cite{TSHJ2020}, and demonstrate that it is possible to use the embedding for graph machine learning tasks throught implementing a quantum classifier on the top of it. However, their algorithm cannot be used directly to deal with the case where the element of $W$ is a Gaussian similarity function. In $2022$, Li et al. designed a Laplacian eigenmap algorithm based on variational quantum generalized eigensolver \cite{LMYZ2022} and their simulation results demonstrate that the proposed algorithm has good convergence. However, their algorithm employs the controlled SWAP test and maximum searching algorithm \cite{chen2020} to construct $W$, which cannot be implemented on NISQ devices. In addition, how to design a good strategy to suppress the barren plateau phenomenon in variational quantum algorithms \cite{MJBS2018} is also a thorny problem. Therefore, designing a variational quantum algorithm that can solve the above problems to solve eigenproblem of $L$ may be an important direction of future work.

\begin{acknowledgments}
We thank Shanwei Ma for useful discussions on the subject. This work is supported by the Beijing Natural Science Foundation (Grant No. 4222031), the National Natural Science Foundation of China (Grant Nos. 61976024, 61972048, 62006105), BUPT innovation and entrepreneurship support program (Grant No. 2021-YC-A206).
\end{acknowledgments}

\begin{widetext}
\appendix
\section{A detailed analysis of the last inequality of Eq.$(17)$ and Eq.$(34)$}
\label{Sec:analysis}

Before we analyze the last inequality of Eq.$(17)$ and Eq.$(34)$ in detail, we first give the lemmas as follows:

$\mathbf{LemmaA1}$ (Error propagation of quantum states)

If $\||\mathbf{x}\rangle-|\mathbf{y}\rangle\|_{2}\leq\varepsilon$, then
\begin{equation}
\|a|\mathbf{x}\rangle-b|\mathbf{y}\rangle\|_{2}\leq(a-b)\||\mathbf{x}\rangle\|_{2}+b\||\mathbf{x}\rangle-|\mathbf{y}\rangle\|_{2}\leq a-b+b\varepsilon,
\end{equation}
where $a,b$ are any positive real numbers.

Proof:
\begin{equation}
\begin{aligned}
\|a|\mathbf{x}\rangle-b|\mathbf{y}\rangle\|_{2}&=\|a|\mathbf{x}\rangle-b|\mathbf{x}\rangle+b|\mathbf{x}\rangle- b|\mathbf{y}\rangle\|_{2}\leq\|(a-b)|\mathbf{x}\rangle\|_{2}+\|b(|\mathbf{x}\rangle-|\mathbf{y}\rangle)\|_{2}\\
&\leq(a-b)\||\mathbf{x}\rangle\|_{2}+b\||\mathbf{x}\rangle-|\mathbf{y}\rangle\|_{2}\leq a-b+b\varepsilon.
\end{aligned}
\end{equation}

$\mathbf{LemmaA2}$ (Error propagation of tensor products of quantum states)

If $\||\mathbf{x}\rangle-|\mathbf{y}\rangle\|_{2}\leq\varepsilon$, then $\|| \mathbf{x}\rangle^{\otimes p}-|\mathbf{y}\rangle^{\otimes p}\|_{2}\leq p\varepsilon$, where $p$ is any positive real numbers.

Proof: we prove it by mathematical induction.

When $p=2$, we have
\begin{equation}
\begin{aligned}
\||\mathbf{x}\rangle^{\otimes 2}-|\mathbf{y}\rangle^{\otimes 2}\|_{2}&=\|| \mathbf{x}\rangle\otimes(|\mathbf{x}\rangle-|\mathbf{y}\rangle)+(| \mathbf{x}\rangle-|\mathbf{y}\rangle)\otimes|\mathbf{y}\rangle\|_{2}\leq\||\mathbf{x}\rangle\|_{2}\||\mathbf{x}\rangle-|\mathbf{y}\rangle\|_{2}+\||\mathbf{x}\rangle-|\mathbf{y}\rangle\|_{2}\||\mathbf{y}\rangle\|_{2}\\
&=\||\mathbf{x}\rangle-|\mathbf{y}\rangle\|_{2}(\||\mathbf{x}\rangle\|_{2}+\||\mathbf{y}\rangle\|_{2})\leq2\varepsilon.
\end{aligned}
\end{equation}
Assume that when $k=p-1$, we get
$\||\mathbf{x}\rangle^{\otimes(p-1)}-|\mathbf{y}\rangle^{\otimes(p-1)}\|_{2}\leq(p-1)\varepsilon$.

When $k=p$, we obtain
\begin{equation}
\begin{aligned}
&\|| \mathbf{x}\rangle^{\otimes p}-|\mathbf{y}\rangle^{\otimes p}\|_{2}=\|| \mathbf{x}\rangle^{\otimes (p-1)}\otimes(|\mathbf{x}\rangle-|\mathbf{y}\rangle)+(|\mathbf{x}\rangle^{\otimes (p-1)}-|\mathbf{y}\rangle^{\otimes (p-1)})\otimes|\mathbf{y}\rangle\|_{2}\\
&=\||\mathbf{x}\rangle^{\otimes (p-1)}\|_{2}\|(|\mathbf{x}\rangle-|\mathbf{y}\rangle)\|_{2}+\|(|\mathbf{x}\rangle^{\otimes (p-1)}-|\mathbf{y}\rangle^{\otimes (p-1)})\|_{2}\||\mathbf{y}\rangle\|_{2}\leq\varepsilon+(p-1)\varepsilon=p\varepsilon.
\end{aligned}
\end{equation}
Then we use the lemmas above to analyze the last inequality of Eq.$(17)$ and Eq.$(34)$. At the same time, we expect that the lemmas above can be used to deal with the error analysis of other quantum algorithms in the future.

(i) The last inequality of Eq.$(17)$:
\begin{equation}
\begin{aligned}
&\||\hat{\Phi}(\mathbf{x}_{i})\rangle-|\Phi(\mathbf{x}_{i})\rangle\|_{2}=\|\frac{1}{\sqrt{\hat{a}}}\sum_{k=0}^{p}\sqrt{\hat{a}_{k}}|k\rangle|0\rangle^{\otimes p-k}|\mathbf{\hat{x}}_{i}\rangle^{\otimes k}-\frac{1}{\sqrt{\tilde{a}}}\sum_{k=0}^{p}\sqrt{\tilde{a}_{k}}|k\rangle|0\rangle^{\otimes p-k}|\mathbf{\tilde{x}}_{i}\rangle^{\otimes k}\|_{2}\\
&\leq(\frac{1}{\sqrt{\hat{a}}}-\frac{1}{\sqrt{\tilde{a}}})\|\sum_{k=0}^{p}\sqrt{\hat{a}_{k}}|k\rangle|0\rangle^{\otimes p-k}|\mathbf{\hat{x}}_{i}\rangle^{\otimes k}\|_{2}+\frac{1}{\sqrt{\tilde{a}}}\|\sum_{k=0}^{p}\sqrt{\hat{a}_{k}}|k\rangle|0\rangle^{\otimes p-k}|\mathbf{\hat{x}}_{i}\rangle^{\otimes k}-\sum_{k=0}^{p}\sqrt{\tilde{a}_{k}}|k\rangle|0\rangle^{\otimes p-k}|\mathbf{\tilde{x}}_{i}\rangle^{\otimes k}\|_{2}\\
&=s+\frac{1}{\sqrt{\tilde{a}}}\|(\sqrt{\hat{a}_{0}}-\sqrt{\tilde{a}_{0}})|0\rangle|0\rangle^{\otimes p}+\sum_{k=1}^{p}|k\rangle|0\rangle^{\otimes (p-k)}(\sqrt{\hat{a}_{k}}|\mathbf{\hat{x}}_{i}\rangle^{\otimes
k}-\sqrt{\tilde{a}_{k}}|\mathbf{\tilde{x}}_{i}\rangle^{\otimes
k})\|_{2}\\
&=s+\frac{1}{\sqrt{\tilde{a}}}\|(\sqrt{\hat{a}_{0}}-\sqrt{\tilde{a}_{0}})|0\rangle|0\rangle^{\otimes p}+\sum_{k=1}^{p}|k\rangle|0\rangle^{\otimes (p-k)}[(\sqrt{\hat{a}_{k}}-\sqrt{\tilde{a}_{k}})|\mathbf{\hat{x}}_{i}\rangle^{\otimes
k}+\sqrt{\tilde{a}_{k}}(|\mathbf{\hat{x}}_{i}\rangle^{\otimes
k}-|\mathbf{\tilde{x}}_{i}\rangle^{\otimes
k})]\|_{2}\\
&\leq s+\frac{1}{\sqrt{\tilde{a}}}\|\sum_{k=0}^{p}(\sqrt{\hat{a}_{k}}-\sqrt{\tilde{a}_{k}})|k\rangle|0\rangle^{\otimes(p-k)}|\mathbf{\hat{x}}_{i}\rangle^{\otimes k}\|_{2}+\frac{1}{\sqrt{\tilde{a}}}\|\sum_{k=1}^{p}|k\rangle|0\rangle^{(p-k)}\sqrt{\tilde{a}_{k}}(|\mathbf{\hat{x}}_{i}\rangle^{\otimes k}-|\mathbf{\tilde{x}}_{i}\rangle^{\otimes k})\|_{2}\\
&\leq s+\frac{1}{\sqrt{\tilde{a}}}\sqrt{\sum_{k=0}^{p}(\sqrt{\hat{a}_{k}}-\sqrt{\tilde{a}_{k}})^{2}}+\frac{1}{\sqrt{\tilde{a}}}\sum_{k=1}^{p}\sqrt{\tilde{a}_{k}}\|(|\mathbf{\hat{x}}_{i}\rangle^{\otimes k}-|\mathbf{\tilde{x}}_{i}\rangle^{\otimes k})\|_{2}\\
&\leq\varepsilon_{\tilde{a}}+\frac{1}{\sqrt{\tilde{a}}}(\sqrt{\tilde{a}_{1}}\varepsilon_{x}+\sqrt{\tilde{a}_{2}}2\varepsilon_{x}+\cdots+\sqrt{\tilde{a}_{p}}p\varepsilon_{x})\leq\varepsilon_{\tilde{a}}+(1+2+\cdots+p)\varepsilon_{x},
\end{aligned}
\label{A5}
\end{equation}
where $s=(\frac{1}{\sqrt{\hat{a}}}-\frac{1}{\sqrt{\tilde{a}}})\sqrt{\hat{a}}$ the first inequality comes from applying the $\mathbf{LemmaA1}$ and the penultimate inequality comes from
\begin{equation}
\begin{aligned}
\|\frac{1}{\sqrt{\hat{a}}}\sum_{k=0}^{p}\sqrt{\hat{a}_{k}}|k\rangle-\frac{1}{\sqrt{\tilde{a}}}\sum_{k=0}^{p}\sqrt{\hat{a}_{k}}|k\rangle\|_{2}&\leq(\frac{1}{\sqrt{\hat{a}}}-\frac{1}{\sqrt{\tilde{a}}})\|\sum_{k=0}^{p}\sqrt{\hat{a}_{k}}|k\rangle\|_{2}+\frac{1}{\sqrt{\tilde{a}}}\|\sum_{k=0}^{p}(\sqrt{\hat{a}_{k}}-\sqrt{\tilde{a}_{k}})|k\rangle\|_{2}\\
&\leq(\frac{1}{\sqrt{\hat{a}}}-\frac{1}{\sqrt{\tilde{a}}})\sqrt{\hat{a}}+\frac{1}{\sqrt{\tilde{a}}}\sqrt{\sum_{k=0}^{p}(\sqrt{\hat{a}_{k}}-\sqrt{\tilde{a}_{k}})^{2}}\leq\varepsilon_{\tilde{a}}
\end{aligned}
\end{equation}
and the $\mathbf{LemmaA2}$.

(ii) The last inequality of Eq.$(34)$:
\begin{equation}
\begin{aligned}
&\||\hat{\Psi}(\mathbf{x}_{i})\rangle-|\Psi(\mathbf{x}_{i})\rangle\|_{2}=\|\exp(-\lambda\|\mathbf{x}_{i}\|^{2})(\sum_{k=0}^{p}\sqrt{\hat{a}_{k}}\|\mathbf{x}_{i}\|^{k}|k\rangle|0\rangle^{\otimes(p-k)}|\hat{\mathbf{x}}_{i}\rangle^{\otimes
k}-\sum_{k=0}^{p}\sqrt{a_{k}}\|\mathbf{x}_{i}\|^{k}|k\rangle|0\rangle^{\otimes(p-k)}|\mathbf{x}_{i}\rangle^{\otimes
k})\|_{2}\\
&=\exp(-\lambda\|\mathbf{x}_{i}\|^{2})\|(\sqrt{\hat{a}_{0}}-\sqrt{a_{0}})|0\rangle|0\rangle^{\otimes p}+\sum_{k=1}^{p}|k\rangle|0\rangle^{\otimes p-k}\|\mathbf{x}_{i}\|^{k}[(\sqrt{\hat{a}_{k}}-\sqrt{a_{k}})|\mathbf{\hat{x}}_{i}\rangle^{\otimes k}+\sqrt{a_{k}}(|\mathbf{\hat{x}}_{i}\rangle^{\otimes k}-|\mathbf{x}_{i}\rangle^{\otimes k})]\|_{2}\\
&\leq\|\sum_{k=0}^{p}(\sqrt{\hat{a}_{k}}-\sqrt{a_{k}})|k\rangle|0\rangle^{\otimes p-k}\|\mathbf{x}_{i}\|^{k}|\mathbf{\hat{x}}_{i}\rangle^{\otimes k}\|_{2}+\|\sum_{k=1}^{p}\sqrt{a_{k}}|k\rangle|0\rangle^{\otimes (p-k)}\|\mathbf{x}_{i}\|^{k}(|\mathbf{\hat{x}}_{i}\rangle^{\otimes k}-|\mathbf{x}_{i}\rangle^{\otimes k})\|_{2}\\
&\leq\sqrt{\sum_{k=0}^{p}(\sqrt{\hat{a}_{k}}-\sqrt{a}_{k})^{2}\|\mathbf{x}_{i}\|^{2k}}+\sum_{k=1}^{p}\sqrt{a_{k}}\|\mathbf{x}_{i}\|^{k}\||\mathbf{\hat{x}}_{i}\rangle^{\otimes k}-|\mathbf{x}_{i}\rangle^{\otimes k}\|_{2}\leq\sqrt{\sum_{k=0}^{p}(\sqrt{\hat{a}_{k}}-\sqrt{a}_{k})^{2}\|\mathbf{x}_{i}\|^{2k}}+\sum_{k=1}^{p}\sqrt{a_{k}}\|\mathbf{x}_{i}\|^{k}k\varepsilon_{x},\\
\end{aligned}
\end{equation}
where the last inequality comes from $\mathbf{LemmaA2}$.

When $\|\mathbf{x}_{i}\|>1$, we have
\begin{equation}
\begin{aligned}
\||\hat{\Psi}(\mathbf{x}_{i})\rangle-|\Psi(\mathbf{x}_{i})\rangle\|_{2}&\leq\|\mathbf{x}_{i}\|^{p}\sqrt{\sum_{k=0}^{p}(\sqrt{\hat{a}_{k}}-\sqrt{a}_{k})^{2}}+\|\mathbf{x}_{i}\|^{p}\sqrt{a}\cdot(\frac{1}{\sqrt{a}}\sum_{k=1}^{p}\sqrt{a_{k}}k\varepsilon_{x})\\
&\leq\|\mathbf{x}_{i}\|^{p}[\sqrt{a}\varepsilon_{a}+\sqrt{a}\sum_{k=1}^{p}k\varepsilon_{x}]=\sqrt{a}\|\mathbf{x}_{i}\|^{p}[\varepsilon_{a}+(1+2+\cdots+p)\varepsilon_{x}],\\
\end{aligned}
\end{equation}
where the second inequality comes from the Eq.$(A5)$.

When $0<\|\mathbf{x}_{i}\|<1$, we get
\begin{equation}
\||\hat{\Psi}(\mathbf{x}_{i})\rangle-|\Psi(\mathbf{x}_{i})\rangle\|_{2}\leq\sqrt{\sum_{k=0}^{p}(\sqrt{\hat{a}_{k}}-\sqrt{a}_{k})^{2}}+\sqrt{a}\cdot(\frac{1}{\sqrt{a}}\sum_{k=1}^{p}\sqrt{a_{k}}k\varepsilon_{x})\leq\sqrt{a}[\varepsilon_{a}+(1+2+\cdots+p)\varepsilon_{x}].
\end{equation}

\section{The quantum circuit of function $f(x)=\exp(-\lambda x)$}
\label{Sec:model}

 The Taylor expansion of $f(x)=\exp(-\lambda x)$ is shown as:
\begin{equation}
\begin{aligned}
\exp(-\lambda x)&=1-\lambda x+\frac{(\lambda x)^{2}}{2!}+\cdots+\frac{(-1)^{k}(\lambda x)^{k}}{k!}+\frac{(-1)^{k+1}(\lambda\xi)^{k+1}}{(k+1)!},\xi\in(0,x).
\end{aligned}
\end{equation}
According to Taylor's theorem \cite{KM1998}, we know that the $(k+1)$th term in the expansion is $[(-1)^{k+1}(\lambda\xi)^{k+1}]/[(k+1)!]$ and the derivative of $f(x)$ are bounded. We can design the quantum circuit of $f(x)$ by the Quantum Multiply-Adder (QMA) \cite{ZSS2017}, which is shown in Fig.~\ref{FIG:5}, with $O(\poly\log(1/\epsilon))$ one- or two- qubits gates, where $\epsilon$ is the accuracy of the algorithm.
\begin{figure}
\centering
	\includegraphics[width=15cm]{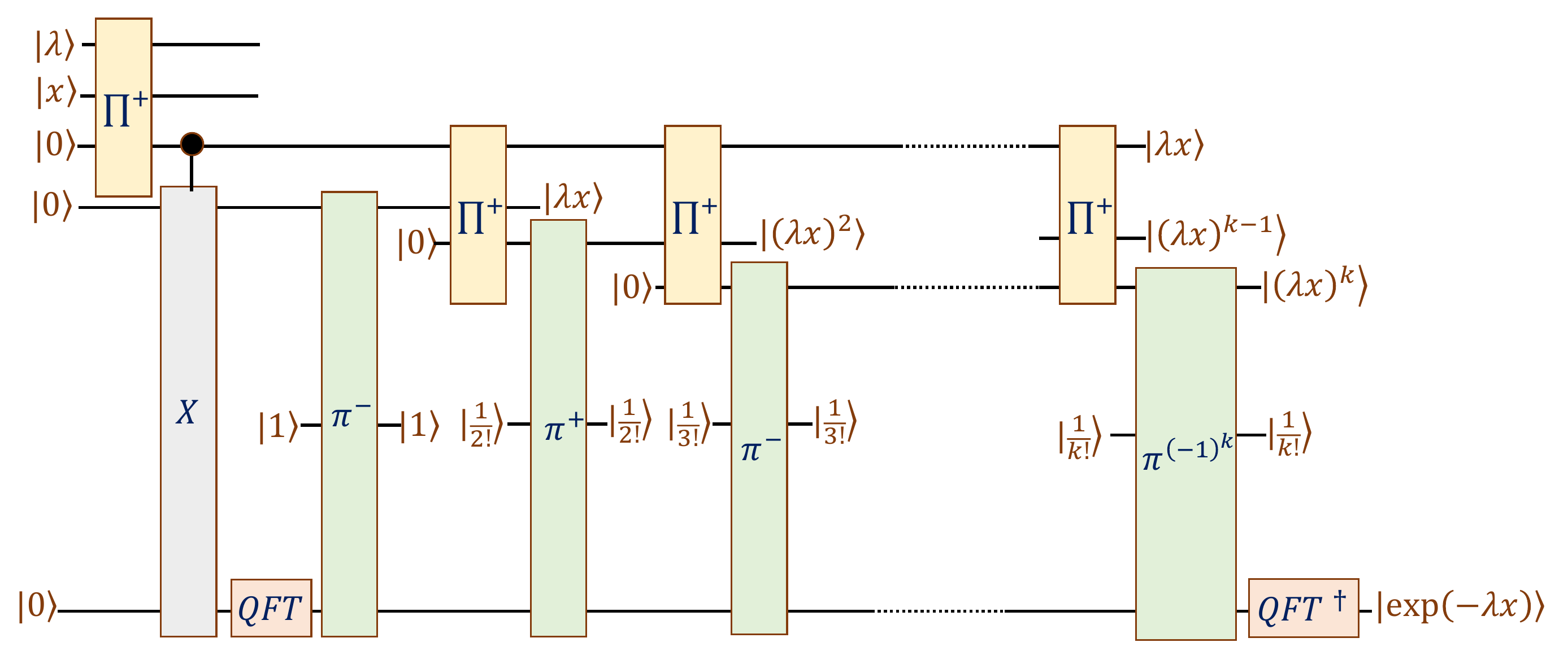}
	\caption{The quantum circuit of $f(x)=\exp(-\lambda x)$. The controlled operator $X$ denote the Pauli operator $X$. $QFT$ and $QFT^{\dagger}$ stand for the quantum Fourier transformation and the inverse quantum Fourier transformation, respectively. The operator $\Pi^{+}$ can implement the following transformation: $\Pi^{+}|\lambda\rangle|x\rangle|0\rangle=|\lambda\rangle|x\rangle|\lambda x\rangle$. It can be represented as $\Pi^{+}=(I\otimes I \otimes QFT^{\dagger})\cdot \pi^{\pm}\cdot(I\otimes I \otimes QFT)$, where $\pi^{\pm}|a\rangle|b\rangle|\phi(c)\rangle=|a\rangle|b\rangle|\phi(c\pm a\cdot b)\rangle$ with $|\phi(c)\rangle=QFT|c\rangle$ and $a,b,c$ are the input qubits.}
	\label{FIG:5}
\end{figure}

\section{Implement the quantum simulation of $L$ in the case of $\|\mathbf{x}_{i}\|=1,i=1,\cdots,n$ }
\label{Sec:CI}

In this section, we analyze the quantum simulation of implementing $L$ in the case of $\|\mathbf{x}_{i}\|=1,i=1,\cdots,n$. According to Eq.~\ref{16} in Sec.~\ref{Sec:W}, we can obtain $W=\tilde{a}(n\rho_{0}-I)$. Due to ${\rm Tr}(L)={\rm Tr}(D)$, we have
\begin{equation}
\bar{L}=\frac{L}{{\rm Tr}(L)}=\frac{D-W}{{\rm Tr}(L)}=\frac{D}{{\rm Tr}(D)}-\frac{n\tilde{a}}{{\rm Tr}(D)}\rho_{0}+\frac{I}{{\rm Tr}(I)}\frac{\tilde{a}{\rm Tr}(I)}{{\rm Tr}(D)}:=\rho_{1}-d\rho_{0}+e\rho_{3},
\end{equation}
where $d=n\tilde{a}/{\rm Tr}(D),e=\tilde{a}{\rm Tr}(I)/{\rm Tr}(D)$.

Similarly to Sec.~\ref{Sec:D}, we can implement the quantum simulation of $\bar{L}$.

\section{Reveal the eigeninformation of $W$}
\label{Sec:Re}

In this section, we introduce that the algorithm to reveal the eigeninformation of $W$ as follows:

For the case of $\|\mathbf{x}_{i}\|=1$, we can obtain $W=\tilde{a}(n\rho_{0}-I)$. Thus, we have
\begin{equation}
W/n=\tilde{a}(\rho_{0}-I/n)=\tilde{a}\rho_{0}-\tilde{a}\rho_{3}.
\label{D.1}
\end{equation}

For the case of $|\mathbf{x}_{i}\|\neq1$, we can obtain $W={\rm Tr}(W+I)\rho_{1}-I$. Due to ${\rm Tr}(W+I)=n$, Thus, we have
\begin{equation}
W/n=\rho_{1}-\rho_{3}.
\label{D.2}
\end{equation}

 In short, Eq.~\ref{D.1} and Eq.~\ref{D.2} can be viewed as a linear combination of block-encoded operators, respectively. Similarly to Sec.~\ref{Sec:D}, we can implement the quantum simulation of $W$. Then we can solve the eigenproblem of $W$ by using the algorithm in Sec.~\ref{Sec:Extract}.

\end{widetext}

\bibliography{biblatex-phys}
\end{document}